\documentclass[10pt,onecolumn,letterpaper]{article}
\usepackage{graphicx}
\usepackage{comment}
\usepackage{amsmath,amssymb} 
\usepackage{color}
\usepackage{hyperref}
\usepackage{booktabs}

\usepackage[width=122mm,left=12mm,paperwidth=146mm,height=193mm,top=12mm,paperheight=217mm]{geometry}
\providecommand{\keywords}[1]{\textbf{\textit{Keywords:}} #1}

\begin{document}

\title{DeepHist: Differentiable Joint and Color Histogram Layers for Image-to-Image Translation} 
\author{
\normalsize
\textit{Mor Avi-Aharon, Assaf Arbelle, and Tammy Riklin~Raviv }
}
\date{
\normalsize
Department of Electrical and Computer Engineering
\\ Ben-Gurion University of the Negev, Beer-Sheva, Israel
}

\maketitle

\begin{abstract}
We present the DeepHist - a novel Deep Learning framework for augmenting a network by histogram layers and demonstrate its strength by addressing image-to-image translation problems. Specifically, given an input image and a reference color distribution we aim to generate an output image with the structural appearance (content) of the input (source) yet with the colors of the reference. 
The key idea is a new technique for a differentiable construction of joint and color histograms of the output images.
We further define a color distribution loss based on the Earth Mover's Distance between the output’s and the reference’s color histograms and a Mutual Information loss based on the joint histograms of the source and the output images.
Promising results are shown for the tasks of color transfer, image colorization and edges $\rightarrow$ photo, where the color distribution of the output image is controlled.
Comparison to Pix2Pix and CyclyGANs are shown.
\end{abstract}

\section{Introduction}
Convolutional Neural Networks (CNNs) dramatically improve the state-of-the-art in many practical domains~\cite{krizhevsky2012imagenet,lecun1998gradient}.
While numerous loss functions were proposed, metrics based on image histograms, which represent images by their color distributions~\cite{Gonzalez2008,Szeliski2010} are not considered.
The main obstacle seems to be the histogram construction which is not a differentiable operation and therefore cannot be incorporated into a deep learning framework.

\begin{figure}[t]
\begin{center}
\includegraphics[width=\linewidth]{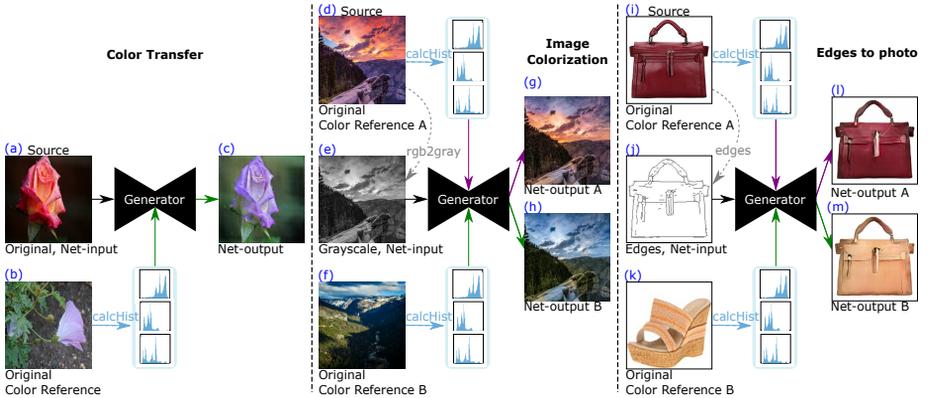}
\end{center}
\caption{Image-to-image translation tasks are presented from left to right: color transfer, image colorization and edges$\rightarrow$photo. 
The inputs for all tasks consist of a content reference image 
(an edge map in the case of edge$\rightarrow$photo) and the color histograms of an RGB image (a color-reference image).
The outputs for all the tasks is an RGB image with the content of the source image and the color distributions of the color-reference image.
For example, (l) and (m) are two possible outputs of the edges$\rightarrow$photo, for the input histogram of either (i) or (k), respectively.
}
\label{fig:tasks}
\end{figure}

In this work, we introduce the DeepHist - a deep learning framework for image generation, which enables a differentiable construction of joint and color histograms of the output images. We further define color-based and statistical similarity loss functions that are exclusively built on the differentiable histograms of the generated images. Specifically, we augment a neural network generator by histogram layers that take part in the back-propagation process in which the respective histogram loss functions are used for updating the generator weights. 
Relying on the color distribution rather than on the differences between corresponding pixels allows us to address image-to-image translation problems for which the desired, target images do not necessarily exist. Consider for example the color transfer problem, as exemplified in the left panel of Fig.~\ref{fig:tasks}, where the aim is to paint a input (source) image with the colors of a different color reference image. For this kind of unpaired learning tasks, neither of the prevalent loss functions that are based on pixel-by-pixel comparison, e.g., mean-square error (MSE) or cross-entropy, can be used.
We also address generalization of the image colorization and edge$\rightarrow$photo problems, where the color distribution of a generated image is constrained to fit a particular color histogram (Fig.~\ref{fig:tasks} middle and right panels). 

Color and intensity histograms are useful representations for image-to-image translation tasks.
Classical methods for color transfer were based on the concept of histogram matching, where the main idea was to adapt a color histogram of a given image to the target image. Reinhard et al.~\cite{reinhard2001color} addressed color transfer by using a simple statistical
analysis to impose one image’s color characteristics on another, in the Lab color space. Neumann et al.~\cite{neumann2005color} used 3D histogram matching in the hue-saturation-lightness (HSL) colorspace. Their method is based on mapping an arbitrary source gamut to the arbitrary target one, while colors with same hues of target image will have the same hues after the transformation. The proposed mapping required histogram smoothing to reduce undesired gradient effects.

In this work, we exploit histogram matching using the network as an optimizer. The distance between a pair of histograms is defined by the Earth Mover's Distance (EMD).

A deformation of the color distribution of an image can distort its content, therefore enforcement of the structural similarity between the source and the output images is required. The main problem is that images have different intensities in corresponding locations making pixel-to-pixel comparison  not applicable. To address this issue, we suggest to use the mutual information (MI)  of the source and the output images as a measure of their content-based, color-free similarity.
In a seminal work Viola and Wells~\cite{viola1997alignment} used a cost function based on MI for image registration, where the target image and the source have different intensity distributions.
Since then, MI-based registration became popular in biomedical imaging applications, in particular when the alignment of medical images acquired by different imaging modalities is addressed.
An essential component for calculating the MI of two images is the generation of their joint histogram.
In the context of image registration it is called a co-occurrence matrix. While there has been significant work exploiting co-occurrence matrices, the use of joint histograms and MI for image-to-image translation tasks (to the best of our knowledge) has not been done before.
Moreover, differential construction of intensity histograms and joint histograms as part of a deep learning framework is done here for the first time.

Recent image generation approaches and image-to-image translation, in particular are mostly based on deep learning frameworks.
Since the main aim is generating realistic examples, adversarial frameworks, in which an adversarial network is trained on discriminating between real and fake examples, seem to be very effective~\cite{goodfellow2014generative}.
In their pix2pix framework, Isola et al. performed image-to-image translation (e.g., colorization of gray scale images and edges$\rightarrow$photo) by using adversarial loss as well as $L_1$ loss between corresponding pixels in the network's output and the desired target image~\cite{IsolaZZE16}.
In this sense, the pix2pix is a fully supervised method and obviously cannot be applied to problems (such as color transfer) where the desired target image does not exist. Moreover, as discussed in~\cite{IsolaZZE16} the images generated by using L1 loss tend to have grayish or brownish colors when there is an uncertainty regarding to which of several plausible color values a pixel should take on.
Specially, L1 will be minimized by choosing the median of the conditional probability density function over possible colors.
The problem of color-uncertainty is addressed in Zhang et al.~\cite{zhang2016colorful} by a class-based colorization approach, in which the loss of each pixel in an image of a particular class is weighted  the frequency of its color in that class.
This process, termed as class-rebalancing increases the color diversity of the test results.
Zhu et al.~\cite{zhu2017unpaired} referred to image-to-image translation in unpaired setting using cycle-consistent adversarial networks. The Cycle GAN enables style and color transfer (e.g., summer to winter) when the desired output image cannot be used for training. The main idea is using an adversarial loss to map an image X into Y and then mapping Y into X such that the cycle consistency is preserved.  The cycle GAN presents compelling results, yet since in many cases the cyclic consistency constrain is not sufficient, additional supervision and loss functions are often required.
He et al.~\cite{he2019progressive} proposed two-step pipeline for color transfer based on deep semantic correspondences (via VGG19) between an input and a reference images followed by local color transfer in the image domain.  The method provides visually appealing results yet requires  structural and semantic similarity of the reference with respect to the input image. Moreover, the output color distribution can be only controlled by the reference image.

\begin{figure}[t]
\begin{center}
\begin{tabular}{ccccc}
{\scriptsize Input}
&{\scriptsize Source}
&{\scriptsize Output 1}
&{\scriptsize Output 2}
&{\scriptsize Output 3}
\\
\includegraphics[width=0.13\linewidth]{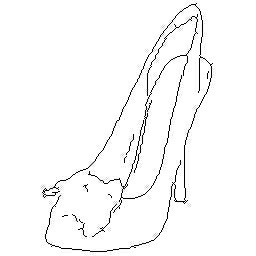}
&\includegraphics[width=0.13\linewidth]{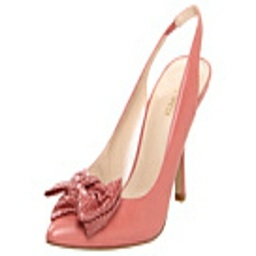}
&\includegraphics[width=0.13\linewidth]{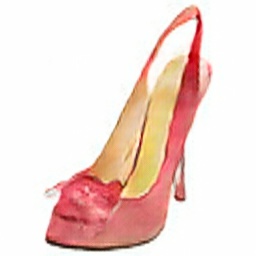}
&\includegraphics[width=0.13\linewidth]{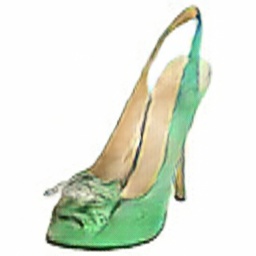}
&\includegraphics[width=0.13\linewidth]{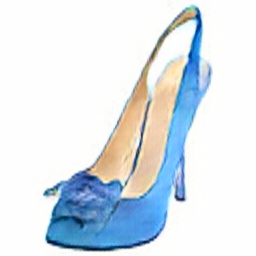}
\\
\includegraphics[width=0.13\linewidth]{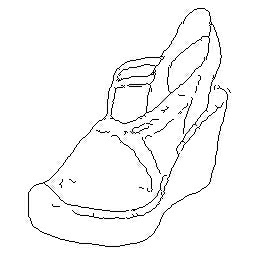}
&\includegraphics[width=0.13\linewidth]{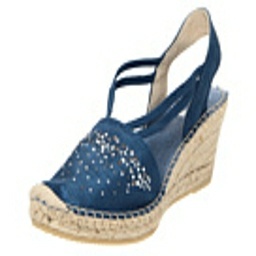}
&\includegraphics[width=0.13\linewidth]{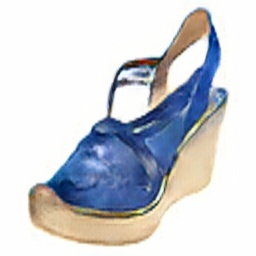}
&\includegraphics[width=0.13\linewidth]{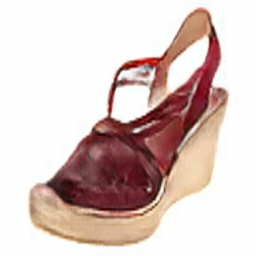}
&\includegraphics[width=0.13\linewidth]{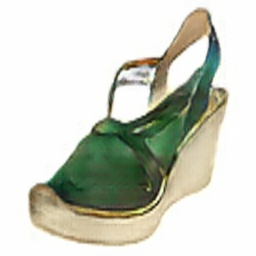}
\end{tabular}
\end{center}
\caption{Edges$\rightarrow$photo based on the same edge image yet with different user-selected color histograms.
The reference color histogram for output 1 is the color histogram of the source image. For output 2-3, we directly defined the color histogram in order to generate images with new colors.}
\label{fig:shoes}
\end{figure}

The DeepHist presents a conceptual alternative to existing image-to-image translation methods. It does not require the extraction of semantic features neither does it need a reference color image with semantic similarity to the input image. Instead, reference color histograms representing the desired color distribution of the output are provided to the network. 
While these color histograms can be constructed from a reference color image - as is the case for the color transfer problem and as exemplified in Fig.~\ref{fig:tasks}, they can be also user-defined for color-controlled image colorization or edges$\rightarrow$photo tasks (Fig.~\ref{fig:shoes}).
The intensity-based loss we propose for `painting' the output image with the reference colors is based on the EMD between a differentiable histogram constructed from the output image and the reference histogram. 
Moreover, structure/content similarity between the source and the output images is preserved thanks to the mutual information loss, which we define based on the joint source-output histogram.
While here as well adversarial loss is utilized for generating realistic images ensuring, for example, green grass and blue sky and not the other way around, our framework does not exclusively or mainly relay on it - making it much more stable. Finally, avoiding the use of pixel-to-pixels comparison via L1 or other distance measures, allows us to handle unpaired image-to-image translation  such as color transfer.

The DeepHist framework is comprised of a generator, which is an adaptation of the well known U-Net - an encoder-decoder with skip connections~\cite{ronneberger2015u}. Yet, the main contributions are the augmented parts of the network which allow differential construction of intensity (1D) and joint (2D) histograms, such that histogram-based loss functions are used to train the image generator in an end-to-end manner.
We demonstrate the proposed frameworks for different paired and unpaired image-to-image translation with several publicly available datasets.
This includes color transfer for the flowers dataset~\cite{Nilsback08}, image colorization for the summer-winter dataset~\cite{zhu2017unpaired} and edges$\rightarrow$photo for the shoes~\cite{yu2014fine} and the bags~\cite{zhu2016generative} datasets.

\section{Methods}
In this section we review the main principles underlying the differentiable construction of 1D and 2D (joint) color histograms (Section~\ref{ssec:DifferentiableHistogramsConstruction}).
We then define the histogram-based metrics  (Section \ref{ssec:Metrics}) that are used for defining the differentiable loss functions (Section~\ref{ssec:Hue-NetLosses}).
The network architecture is presented in Section~\ref{ssec:NetworkArchitecture}.
Implementation details  are presented in Section~\ref{ssec:ImplementationDetails}.

\subsection{Differentiable Histograms Construction}\label{ssec:DifferentiableHistogramsConstruction}

\subsubsection{Color Space}
To address image-to-image translation problems we choose the YUV color space. It is composed of one luma component (Y) and two chrominance components, called U (blue projection) and V (red projection).
The Y channel is in the range $[0,1]$ while the range of the U and the V channels is $[-.5,.5].$
For practical reasons we map all channels' values to $[-1,1].$
In the following Sections we refer to each color channel as a gray-level image.

\subsubsection{Differentiable 1D Color Histogram Formulation}\label{sssec:DifferentiableColorHistogramFormulation}
Images acquired by digital cameras have three color channels each with a discrete range of $K$ intensity values. The intensity distribution of each channel can be described with an intensity histogram obtained by counting the number of pixels in each intensity value.
Considering synthesized images that can take any value in the continuous range $[-1,1],$
we define the intensity of an image pixel $x \in \Omega, $ in a particular channel as $I(x)\in[-1,1].$
We use the Kernel Density Estimation (KDE) for estimating the gray level density $f_I$ of an image's channel $I$ as follows
\begin{equation}
\hat{f}_I (g) = \frac{1}{NW} \sum_{x\in\Omega} \mathcal{K}\left(\frac{I(x)-g}{B}\right)
\label{eq:IntensityDistribution}
\end{equation}
where $g\in[-1,1]$, $\mathcal{K}(\cdot)$ is the kernel, $B$ is the bandwidth and $N=\lvert\Omega\rvert$ is the number of pixels in the image.
We choose the kernel $\mathcal{K}(\cdot)$ as the derivative of the logistic regression function $\sigma(z)$ as follows
\begin{equation}
\mathcal{K}(z) = \frac{d}{dz} \sigma(z) = \sigma(z)\sigma(-z)
\label{eq:kernel}
\end{equation}
where $\sigma(z)=\frac{1}{1+e^{-z}}$.
We note that Eq.~(\ref{eq:kernel}) is a non-negative real-valued integrable function and satisfies the requirements for a kernel (normalization and symmetry).

\begin{figure}[t]
\begin{center}
\bgroup
\def\arraystretch{1}
\begin{tabular}{ccc}
{\scriptsize (a) Output channel} & 
{\scriptsize (b) Activation functions} & 
{\scriptsize (c) Activation maps} 
\\
\includegraphics[height=0.15\linewidth]{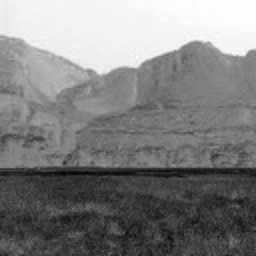}
& \includegraphics[height=0.15\linewidth]{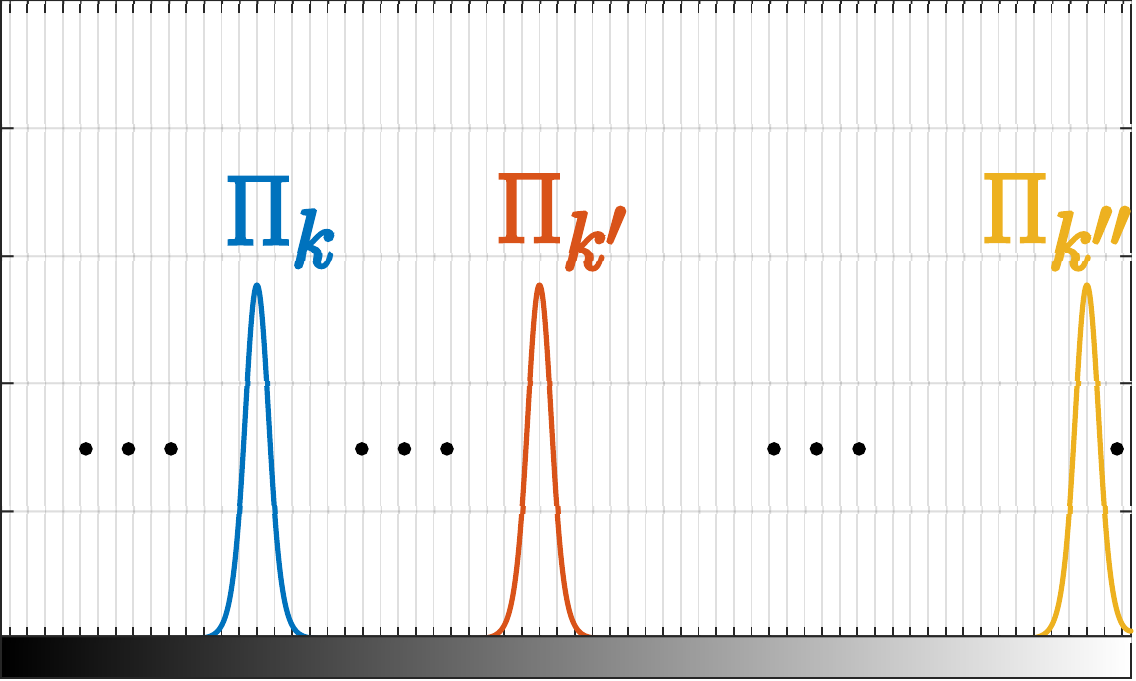}
& \includegraphics[height=0.15\linewidth]{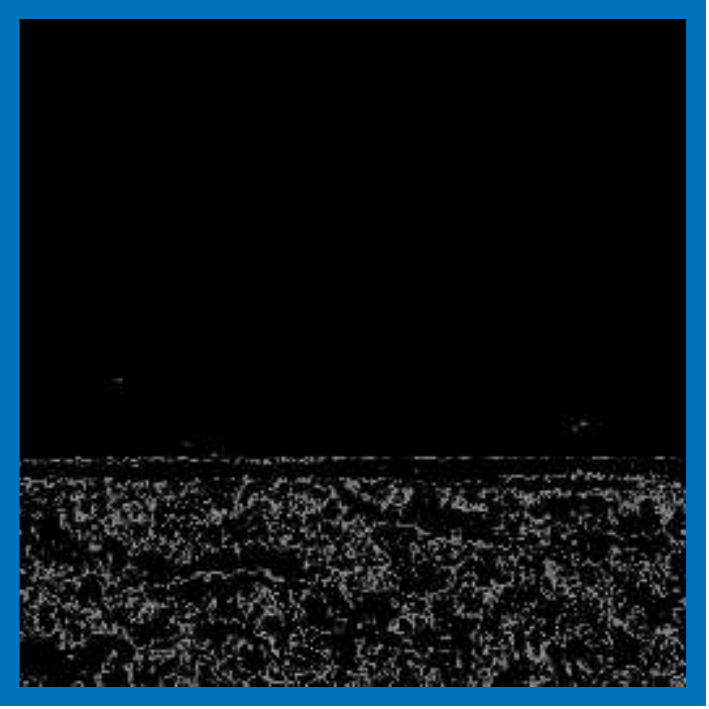}
 \includegraphics[height=0.15\linewidth]{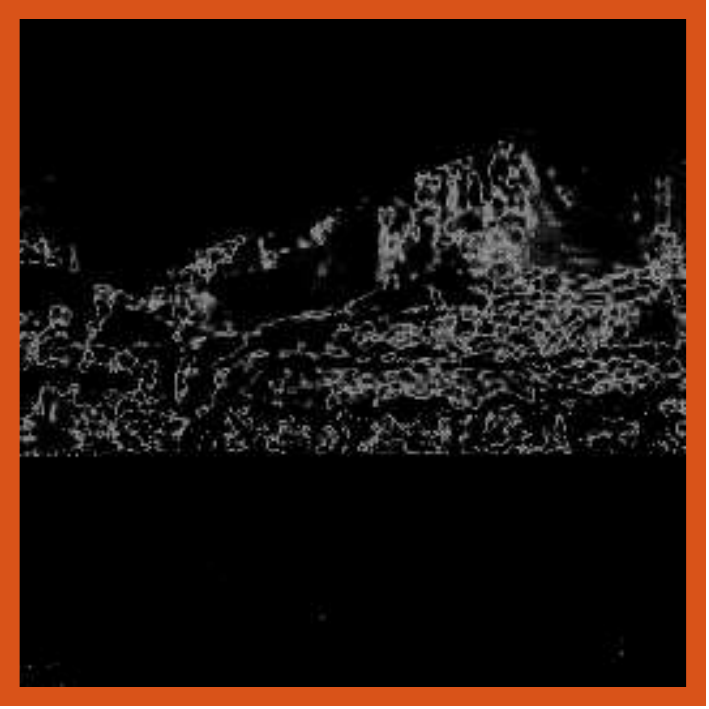}
 \includegraphics[height=0.15\linewidth]{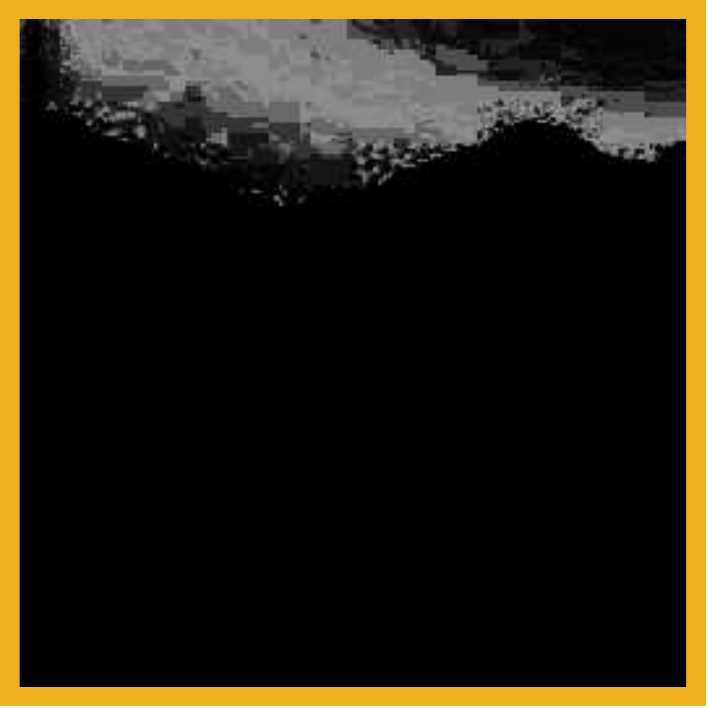}
\end{tabular}
\egroup
\end{center}
\caption{{\bf Activation functions and maps}. (a) One (out of three) color channel of the generated output image. (b) Three out of $K$ activation functions. (c) Three out of $K$ activation maps, each is generated by an application of the respective $\Pi_k$  activation function to the output color channel shown in (a). Note that pixels with values closer to $k$ have higher values in the $k^{th}$ activation map.
}
\label{fig:activation}
\end{figure}

For the construction of smooth and differentiable image histogram, we partition the interval $[-1,1]$ into $K$ sub intervals $\{B_k\}_{k=0}^{K-1}$,
each interval with length $L=\frac{2}{K}$ and center $\mu_k = -1 + L(k+\frac{1}{2})$, then $B_k=[-1+kL,-1+(k+1)L]$.
We then can define the probability of pixel in the image to belong to certain gray level interval (the value of normalized histogram's bin) as
\begin{equation}
P_I(k)  \triangleq \Pr(g\in B_k) = \int_{B_k}{\hat{f}_I(g)dg}
\label{eq:IntensityProbabilityIntegral}
\end{equation}
By solving the integral we get
\begin{equation}
\begin{split}
P_I(k)
&= \frac{1}{N} \sum_{x\in\Omega} \sigma\left(\frac{I(x)-g}{B}\right)\Big|^{\mu_k - L/2}_{\mu_k+L/2} \\
&= \frac{1}{N} \sum_{x\in\Omega} \Bigl[\sigma\left(\frac{I(x)-\mu_k+L/2}{B}\right)
-\sigma\left(\frac{I(x)-\mu_k-L/2}{B}\right)\Bigr]
\end{split}
\label{eq:IntensityProbabilityRect}
\end{equation}
The function $P_I(k)$ which provides the value of the $k^{\mbox{\small{th}}}$ bin in a differentiable histogram
can be rewritten as follows:
\begin{equation}
P_I(k) = \frac{1}{N}\sum_{x\in\Omega} \Pi_k(I(x)),
\label{eq:IntensityProbabilityFinal}
\end{equation}
where,
\begin{equation}
\Pi_k(z) \triangleq \sigma(\frac{z-\mu_k + L/2}{B})-\sigma(\frac{z-\mu_k - L/2}{B})
\label{eq:Pi_k}
\end{equation}
is a differentiable approximation of the Rect function. Fig.~\ref{fig:activation} illustrates the application of three (out of K)  activation functions $\Pi_k$ on a gray scale image. The resulting $K$ channels are used for the construction of the corresponding gray level histogram. Specifically, the $k^{th}$ histogram bin is obtained by a summation of the $k^{th}$ channel values. The set of $K$ channels can be viewed as smooth 1-hot approximations of the pixels values in a gray-level image. Note that the support of $\Pi_k$ is over the gray-level range and as opposed to convolutional kernel it is not spatial.
A differentiable histogram $\mathbf{h}_j$ of a gray-level image $I_j$ is defined as follows:
\begin{equation}
\mathbf{h}_j=\{\mu_{k},P_{I_j}(k)\}_{k=0}^{K-1},\quad j\in\{1,2\}
\label{eq:Histogram}
\end{equation}

\subsubsection{Differentiable Joint Color Histogram Formulation}\label{sssec:Differentiable Joint Color Histogram}
The joint histogram of two gray-level images, each with $K$ discrete gray levels is a $K\times K$ matrix constructed such that its $(k,l)$ entry counts the number of times, pixels with gray level value $k$ in one image correspond to pixels with gray level value $l$ in the other.   The joint gray-level density is obtained by normalizing the joint gray-level histogram.  Considering two images $I_1,I_2:\Omega\rightarrow [-1,1]$ with continues pixel values, their joint gray-level density can be defined using multivariate KDE as follows:
\begin{equation}
\hat{f}_{I_1,I_2} (g_1,g_2) = \frac{1}{N}
\lvert\mathbf{B}\rvert^{-1/2}
\sum_{x\in\Omega}
\mathcal{K} \left( \mathbf{B}^{-1/2} (\mathbf{I}(x) - \mathbf{g}) \right)
\label{eq:JointIntensityDistribution}
\end{equation}
where, $\mathbf{I}(x) = \begin{bmatrix} I_1 & I_2
\end{bmatrix}^T$, $\mathbf{g} = \begin{bmatrix} g_1 & g_2
\end{bmatrix}^T$, $\mathbf{B}$ is the bandwidth (or smoothing) $2\times 2$ matrix and $\mathcal{K}(\cdot,\cdot)$ is the symmetric 2D kernel function.
As in the 1D case (Eq.~\ref{eq:kernel}), we choose the kernel $\mathcal{K}(\cdot,\cdot)$ as the derivative of the logistic regression function $\sigma(z)$ for each of the two variables separately:
\begin{equation}
\mathcal{K}(z_1,z_2) = \frac{d}{dz}\sigma(z_1)  \frac{d}{dz}\sigma(z_2) = \sigma(z_1)\sigma(-z_1)\sigma(z_2)\sigma(-z_2)
\label{Kernel2D}
\end{equation}
We define the bandwidth matrix $\mathbf{B}$ as
$\begin{bmatrix}
B & 0 \\ 0 & B
\end{bmatrix}$.
We define the probability of corresponding pixels in $I_1$ and $I_2$ to belong to the intensity intervals $B_{k_1}$ and $B_{k_2},$ correspondingly, as follows:
\begin{equation}
\begin{split}
P_{I_1,I_2}(k_1,k_2) \triangleq
\Pr(I_1(x)\in B_{k_1}, I_2(x)\in  B_{k_2})
= \int_{B_{k_1}}\int_{B_{k_2}}{\hat{f}_{I_1,I_2}(g_1,g_2)dg_1dg_2}
\end{split}
\label{eq:JointProabilityIntegreal}
\end{equation}
By solving the integral we get:
\begin{equation}
\begin{split}
P_{I_1,I_2}(k_1,k_2) =
\frac{1}{N}\sum_{x\in\Omega}
\sigma\left(\frac{I_1(x)-g_1}{B}\right)
\Big|^{\mu_{k_1} - L/2}_{\mu_{k_1}+L/2}
\times\sigma\left(\frac{I_2(x)-g_2}{B}\right)
\Big|^{\mu_{k_2} - L/2}_{\mu_{k_2}+L/2}
\end{split}
\label{JointProabilityRect}
\end{equation}
By using the definition of $\Pi_k$ from Eq.~\ref{eq:Pi_k}, we can expressed the value of joint histogram $k_1,k_2$-th bin as
\begin{equation}
{P}_{I1,I2}(k_1,k_2) = \frac{1}{{N}}\sum_{x\in\Omega}{\Pi}_{k_1}(I_1(x)){\Pi}_{k_2}(I_2(x))
\label{eq:JointProabilityFinal}
\end{equation}
This equation can be also written using matrix notation.
We define a $K\times N$ matrix $\mathbf{P_j}$ where each of its $K$ rows is a flatten activation map, generated from a gray level image $I_j$.
A differentiable joint histogram $\mathbf{J}$ of two images $I_j$, $j=1,2$ can be constructed via matrix multiplication as follows:
\begin{equation}
\mathbf{J}(I_1,I_2) = \frac{1}{N} \mathbf{P_1} \mathbf{P_2}^T
\label{eq:JointProabilityMatrix}
\end{equation}

\subsubsection{Histogram Layers}\label{sssec:HistogramLayers}

\paragraph{1D Histogram Layer}\label{par:1DHistogram Layer}

\begin{figure}[t!]
\begin{center}
\includegraphics[width=\linewidth]{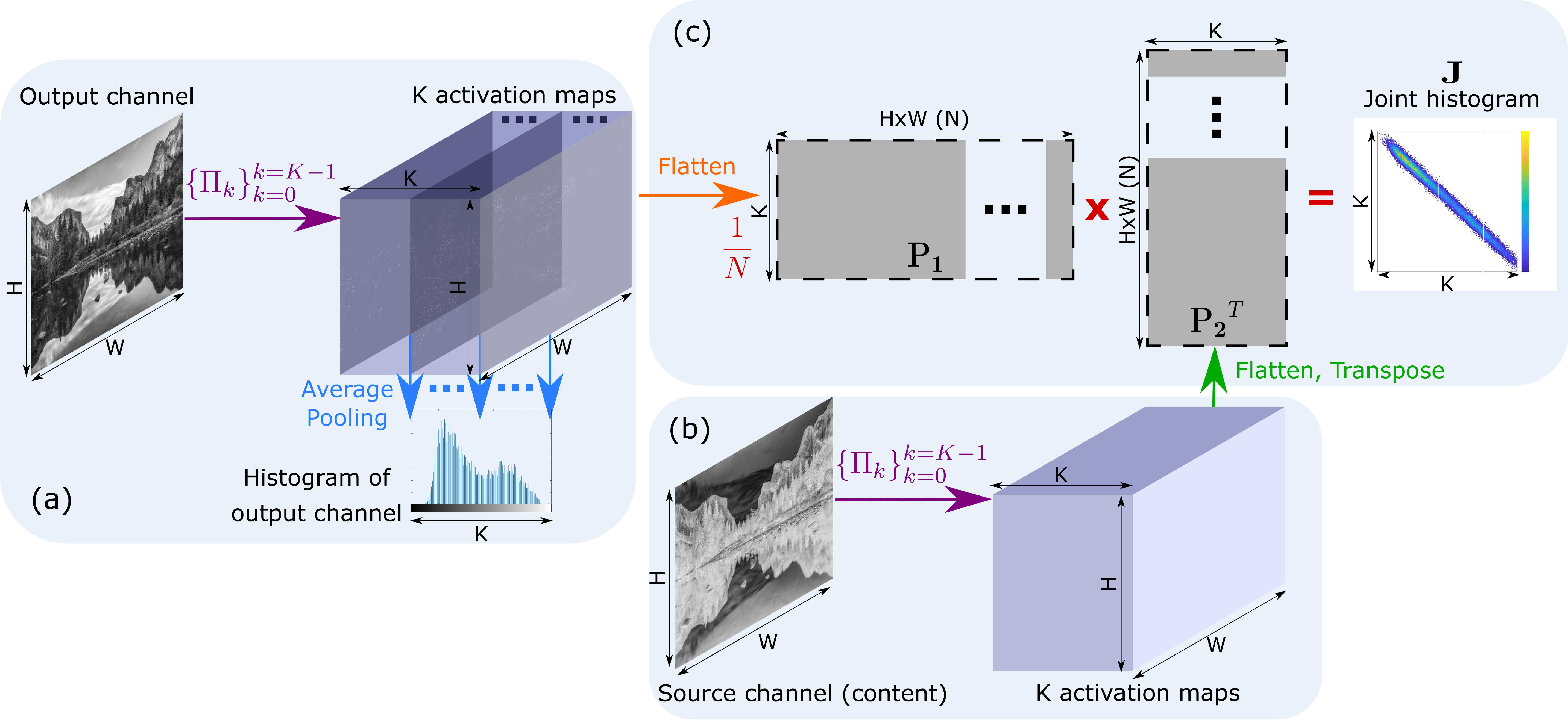}
\end{center}
\caption{\textbf{Histogram construction:}
(a) $K$ activation maps ($H \times W$ matrices) are generated by the application of $K$ activation functions $\{\Pi_k(\cdot)\}_{k=0}^{k=K-1}$ to an output channel of the generator network (one out of three color channels of the synthesized image).  Summation (and normalization by number of pixels $N$) of the values of the $K$ activation maps provides the respective histogram bin.
(b) Construction of $K$ activation maps by the application of $K$ activation functions to each of the color channels of the source image. 
(c) The Joint histogram is constructed by matrix multiplication of the reshaped activation maps ($K \times N$) of the output channel and the reshaped activation maps of the source ($N \times K$). The joint histogram is used for defining the Mutual Information loss to constrain content-based similarity between the generated and the source images.
\label{fig:histogram_layer}}
\end{figure}
Three gray-level (1D) histogram layers of size $K$ are constructed from the output layers of the generator network (the synthesized output image) one for each color channel. The value of the $k^{th}$ unit in an histogram layer is obtained by a summation (and normalization by $N$) of the respective $k^{th}$ activation map (Eq.~\ref{eq:IntensityProbabilityFinal}). As illustrated in Figure~2, the activation maps are constructed by the application of $K$ activation functions to the three output image layers.  This operation is illustrated in Fig.~\ref{fig:histogram_layer}a.

\paragraph{Joint Histogram Layer}\label{par:JointHistogramLayer}
Having $K$ activation maps for each color channel of the synthesized output image, we construct three matrices of size $K\times N,$ by reshaping the $H\times W$ maps into $N\times1$ vectors.
Applying a similar process to the input image, we can now construct three joint histograms via three matrix multiplications (Eq.~\ref{eq:JointProabilityMatrix}), corresponding to the Y,U and V channels.
Figure~\ref{fig:histogram_layer} illustrates the main ideas.

\subsection{Metrics}\label{ssec:Metrics}
\subsubsection{Earth Mover's Distance}\label{sssec:EarthMoversDistance}
We use the EMD~\cite{rubner2000earth}, also known as the  Wasserstein metric~\cite{dobrushin1970prescribing} to define the distance between two image histograms.
Let $\mathbf{h_1}$ and $\mathbf{h_2}$ be the histograms of the images $I_1$ and $I_2$, respectively. We note that when $\mathbf{h_1}$ and $\mathbf{h_2}$ have the same overall mass, the EMD is a true metric~\cite{rubner2000earth}. Moreover, when the compared histograms are also 1D EMD has been shown to be equivalent to Mallows distance, which has a closed-form solution~\cite{ELevina2001}.
Werman et al.~\cite{WERMAN1985} showed that the EMD is equal to the $L_1$ distance between the cumulative histograms.
Following Hou et al.~\cite{LeHou2016} we use the Euclidean distance because it usually converges faster and is easier to optimize with gradient descent~\cite{luenberger1984linear,shalev2011stochastic}:
\begin{equation}
\mathcal{D}_{\scriptsize{\mbox{EMD}}}(\mathbf{h_1}, \mathbf{h_2}) = \sum_{i=0}^{K-1}{\left(\mbox{CDF}_i(\mathbf{h_1})-\mbox{CDF}_i(\mathbf{h_2})\right)^2},
\label{eq:EMD}
\end{equation}
where, $\mbox{CDF}_i(\mathbf{h}_j)$ is the $i$-th element of the cumulative density function of $\mathbf{h_j}$.

\subsubsection{Mutual information}\label{sssec:Mutual information}
The MI of two images $I_1$ and $I_2$ is defined as follows:
\begin{equation}
\begin{aligned}
\mathcal{I}(I_1, I_2) = \sum_{k_1=0}^{K-1} \sum_{k_2=0}^{K-1} P_{I_1,I_2}(k_1,k_2)\log{\frac{P_{I_1,I_2}(k_1, k_2)}{P_{I_1}(k_1)P_{I_2}(k2)}},
\end{aligned}
\label{eq:MI}
\end{equation}
where, $P_{I_1}$ ,$P_{I_2}$ are the image histograms as defined is Eq.~\ref{eq:IntensityProbabilityFinal}, and $P_{I_1,I_2}$ is the joint histogram discussed in Section~\ref{sssec:Differentiable Joint Color Histogram}.
Maximizing the MI between the output and the source image allows us to generate images with color-free statical similarity. Following~\cite{alex2003hierarchical} we define the MI loss as follows:
\begin{equation}
\mathcal{D}_{\scriptsize\mbox{MI}}
(I_1,I_2) = 1 - \frac{\mathcal{I}(I_1,I_2)}
{\mathcal{H}(I_1,I_2)},
\label{eq:D_MI}
\end{equation}
where, $\mathcal{H}(I_1,I_2)$ is the joint entropy of $I_1$, $I_2$ defined as
\begin{equation}
\mathcal{H}(I_1,I_2)=- \sum_{k_1=0}^{K-1} \sum_{k_2=0}^{K-1} P_{I_1,I_2}(k_1,k_2)\log{P_{I_1,I_2}(k_1,k_2)}.
\label{eq:entropy}
\end{equation}
The quantity $\mathcal{D}(I_1,I_2)$ is a metric~\cite{alex2003hierarchical}, with $\mathcal{D}(I_1, I_1) = 0$ and $\mathcal{D}(I_1, I_2) \leq 1$ for all pairs $(I_1,I_2)$. This metric has symmetry, positivity and boundedness properties.

\subsection{Loss functions}\label{ssec:Hue-NetLosses}
The complete loss $\mathcal{L}$ is a weighted sum of three loss functions:
\begin{equation}
\mathcal{L} =
\lambda_{\scriptsize\mbox{EMD}} \mathcal{L}_{\scriptsize\mbox{EMD}}
+ \lambda_{\scriptsize\mbox{MI}} \mathcal{L}_{\scriptsize\mbox{MI}} +
\lambda_{\scriptsize\mbox{ADV}}\mathcal{L}_{\scriptsize\mbox{ADV}}
\label{eq:LossAll}
\end{equation}
where $\mathcal{L}_{\scriptsize\mbox{EMD}}$, $\mathcal{L}_{\scriptsize\mbox{MI}}$, $\mathcal{L}_{\scriptsize\mbox{ADV}}$
are the color loss using EMD, the statistical similarity loss using MI and the adversarial loss, respectively.
The scalars $\lambda_{\scriptsize\mbox{EMD}}$, $\lambda_{\scriptsize\mbox{MI}}$, $\lambda_{\scriptsize\mbox{ADV}}$ are the weights.

The EMD loss is derived from Eq.~\ref{eq:EMD} which defines the EMD between two histograms, the EMD loss between the output and reference color histograms is defined as follows:
\begin{equation}
\mathcal{L}_{\scriptsize\mbox{EMD}} = \frac{1}{3} [
\mathcal{D}_{\scriptsize{\mbox{EMD}}}
(
{\mathbf{h}}_{\scriptsize\mbox{REF}}^{\scriptsize\mbox{Y}},
{\mathbf{h}}_{\scriptsize\mbox{OUT}}^{\scriptsize\mbox{Y}}
)
+
\mathcal{D}_{\scriptsize{\mbox{EMD}}}
(
{\mathbf{h}}_{\scriptsize\mbox{REF}}^{\scriptsize\mbox{U}},
{\mathbf{h}}_{\scriptsize\mbox{OUT}}^{\scriptsize\mbox{U}}
)
+
\mathcal{D}_{\scriptsize{\mbox{EMD}}}
(
{\mathbf{h}}_{\scriptsize\mbox{REF}}^{\scriptsize\mbox{V}},
{\mathbf{h}}_{\scriptsize\mbox{OUT}}^{\scriptsize\mbox{V}}
)
]
\end{equation}
where, $\{\mathbf{h}_{\scriptsize\mbox{REF}}^{\scriptsize\mbox{Y}}, \mathbf{h}_{\scriptsize\mbox{REF}}^{\scriptsize\mbox{U}}, \mathbf{h}_{\scriptsize\mbox{REF}}^{\scriptsize\mbox{V}}\}$,
$\{\mathbf{h}_{\scriptsize\mbox{OUT}}^{\scriptsize\mbox{Y}}, \mathbf{h}_{\scriptsize\mbox{OUT}}^{\scriptsize\mbox{U}}, \mathbf{h}_{\scriptsize\mbox{OUT}}^{\scriptsize\mbox{V}}\}$
are the reference and the output histograms of the YUV channels.

MI loss between the channels of the network's output $\{Y^{\scriptsize{\mbox{OUT}}}, U^{\scriptsize{\mbox{OUT}}}, V^{\scriptsize{\mbox{OUT}}}\}$
and the source image $\{Y^{\scriptsize{\mbox{SRC}}}, U^{\scriptsize{\mbox{SRC}}}, V^{\scriptsize{\mbox{SRC}}}\}$
is based on their relative MI (Eq.~\ref{eq:D_MI}) and defined as follows:
\begin{equation}
\mathcal{L}_{\scriptsize{\mbox{{MI}}}} = \frac{1}{3}[
\mathcal{D}_{\scriptsize{\mbox{MI}}}
(Y^{\scriptsize{\mbox{OUT}}},Y^{\scriptsize{\mbox{SRC}}})
+
\mathcal{D}_{\scriptsize{\mbox{MI}}}
(U^{\scriptsize{\mbox{OUT}}},U^{\scriptsize{\mbox{SRC}}})
+
\mathcal{D}_{\scriptsize{\mbox{MI}}}
(V^{\scriptsize{\mbox{OUT}}},V^{\scriptsize{\mbox{SRC}}})]
\label{eq:LossSemantic}
\end{equation}

We use conditional GAN loss similar~\cite{IsolaZZE16}.
The discriminator learns to distinguish between the output and the source conditioned by the input.
For the color transfer problem, the discriminator input is the source or the output image, without conditioned input.
The objective of the conditional GAN can be expressed as:
\begin{equation}
\arg
\min_G
\max_D
{\mathbb{E}}_{x,y}[\log D(x,y)]+
{\mathbb{E}}_{x,z}[\log(1 - D(x,G(x,z)))]
\end{equation}
where $G$ is the generator, $D$ is the discriminator, $x$ is the input image, $y$ is the output image, and $z$ is noise in the form of dropout.

\subsection{DeepHist Network Architecture}\label{ssec:NetworkArchitecture}
\begin{figure}[t!]
\begin{center}
\includegraphics[width=\linewidth]{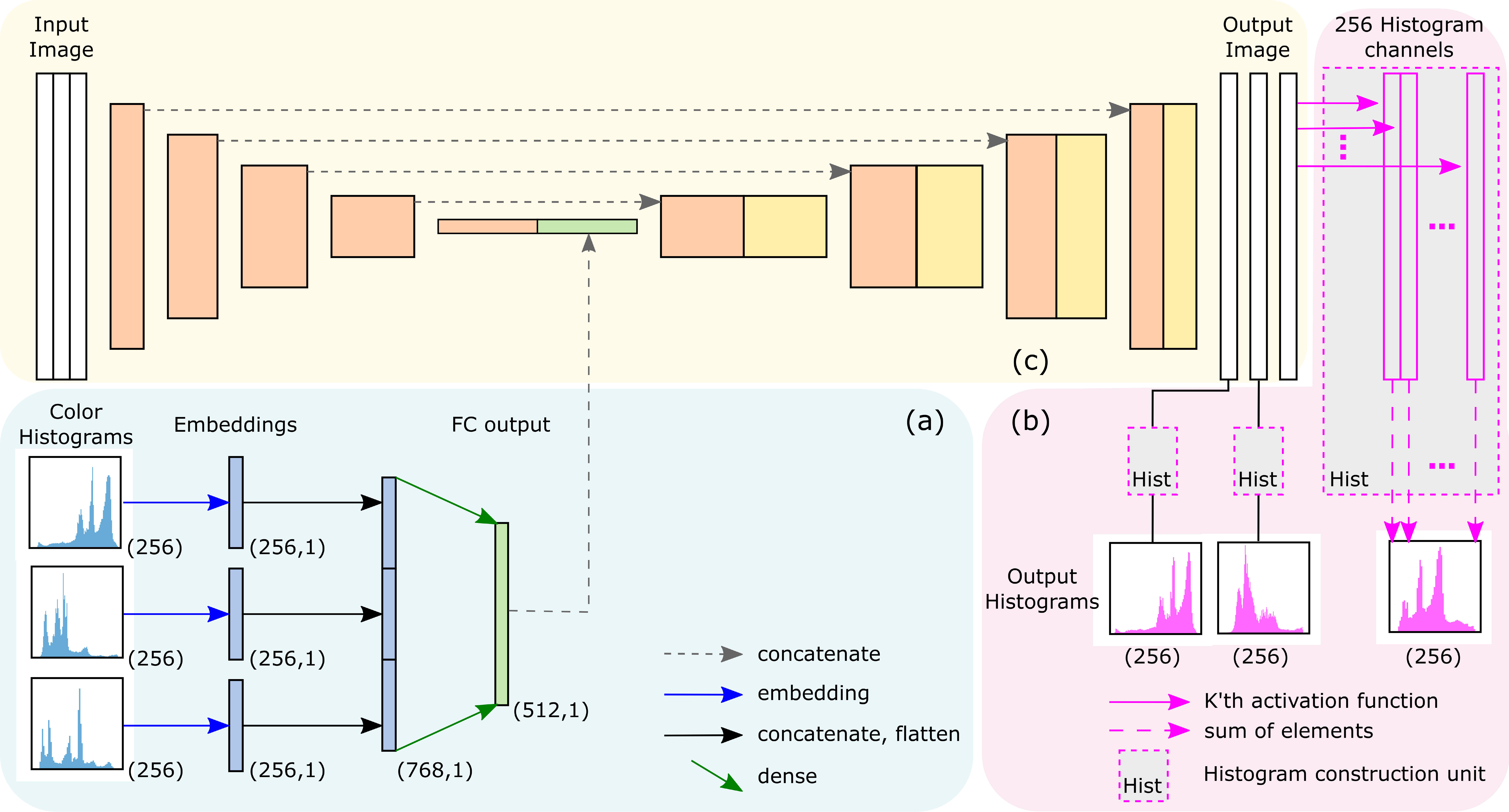}
\end{center}\caption{\textbf{DeepHist network architecture}.
The DeepHist network architecture is composed on an image generator (a modified version of the UNet, yellow color) augmented by input (light blue) and output (pink) histogram layers. The input to the encoder part of the generator is either a gray-scale image (for image colorization), an edge map (for edge$\rightarrow$photo), or a different color images (for color transfer). In addition, target color histograms are fed (each separately) into embedding layers, followed by a fully connected layer and a concatenation with the code layer of the generator.
The three output layers of the generator (which together composed the three color channels of synthesized output image) are used for the construction of color (1D) and joint (2D) histogram layers.  The histograms' construction is illustrated in Figure~\ref{fig:histogram_layer}. \label{fig:Architecture}}
\end{figure}
Figure~\ref{fig:Architecture} illustrates the generator architecture as well as the augmented input and output histogram layers. The DeepHist network architecture is composed on an image generator (a modified version of the UNet~\cite{ronneberger2015u}) augmented by input and output histogram layers. The input to the encoder part of the generator is either a gray-scale image (for image colorization), an edge map (for edge$\rightarrow$photo), or a  RGB image (for color transfer). In addition, reference color histograms are fed (each separately) into embedding layers, followed by a fully connected layer and a concatenation with the code layer of the generator.
Embedding of the reference histogram within the network generator allows us to control the color distribution of the output image.
The three output layers of the generator (which together composed the three color channels of synthesized output image) are used for the construction of color (1D) and joint (2D) histogram layers.  The histogram construction is illustrated in Fig.~\ref{fig:histogram_layer}.
The color histogram layers allow us to constrain color similarity to the reference while the joint histograms layers enable to constrain content similarity to the source via the respective loss functions.
As in~\cite{IsolaZZE16}, we use the convolutional “PatchGAN” classifier~\cite{li2016precomputed} as a discriminator for the construction of an adversarial loss.

\subsection{Implementation Details}\label{ssec:ImplementationDetails}
To optimize our networks, we alternate between one gradient descent step on the Discriminator (D), then one step on the Generator (G).
As suggested in~\cite{goodfellow2014generative}, we train $G$ to maximize $\log{D(x, G(x, z))}$.
We use minibatch SGD and apply the Adam solver~\cite{kingma2014adam}, with a learning rate of $0.0002$, and momentum parameters $\beta_1 = 0.5$, $\beta_2 = 0.999$. 
For histograms construction we use $K=256$ bins, $W=L/2.5$, $L=2/256$.

\section{Experimental Results}\label{sec:ExperimentalResults}
To demonstrate the strengths of DeepHist method, we test
it on a several tasks and datasets:
\begin{enumerate}
\item {\bf Edges $\rightarrow$ photo}
We used two different datasets from~\cite{yu2014fine} and~\cite{zhu2016generative} to demonstrate the edges$\rightarrow$shoe and edges$\rightarrow$bag problems. 
We divided the datasets into training and test as in~\cite{IsolaZZE16}.
During training, the input is an edge map and the output is a synthesized image with the color distribution of the source image (i.e., the real image used for generating the edge map).
The MI loss is calculated with respect to the source image to constrain content similarity to the source.
During the test phase, we generate synthesized images based on the same edge map yet with different selected color histograms. 
For evaluating our method, we present synthesized images with the color distribution of either the source image or a color reference image. Figure~\ref{fig:shoes_table} and Figure~\ref{fig:bags} present visual edges$\rightarrow$shoe and edges$\rightarrow$bag results, respectively.
\begin{figure}
\begin{center}
\begin{tabular}{l}
~~~~~
{\scriptsize Input}\hspace{0.08\linewidth}
{\scriptsize Source}\hspace{0.08\linewidth}
{\scriptsize Our (source)}\hspace{0.04\linewidth}
{\scriptsize Pix2Pix}\hspace{0.04\linewidth}
{\scriptsize Reference}\hspace{0.05\linewidth}
{\scriptsize Our (reference)}
\\
\includegraphics[width=0.15\linewidth]{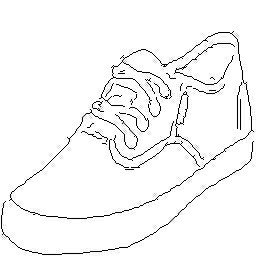}
\includegraphics[width=0.15\linewidth]{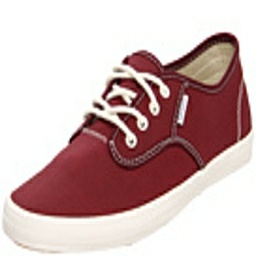}
\includegraphics[width=0.15\linewidth]{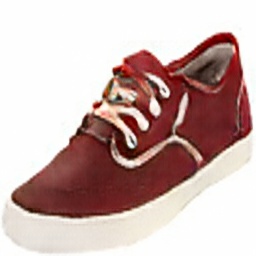}
\includegraphics[width=0.15\linewidth]{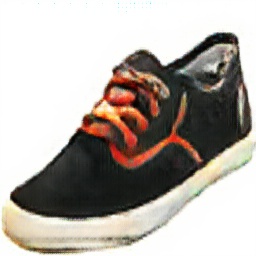}
\includegraphics[width=0.15\linewidth]{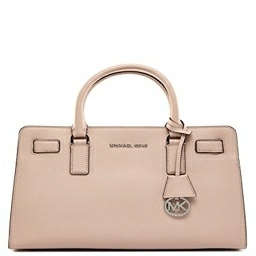}
\includegraphics[width=0.15\linewidth]{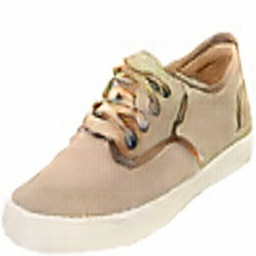}
\\
\includegraphics[width=0.15\linewidth]{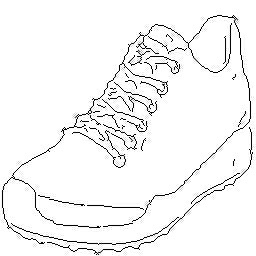}
\includegraphics[width=0.15\linewidth]{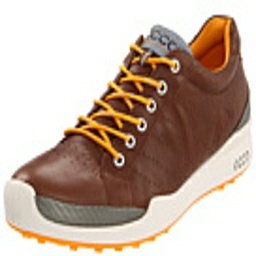}
\includegraphics[width=0.15\linewidth]{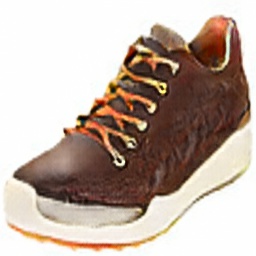}
\includegraphics[width=0.15\linewidth]{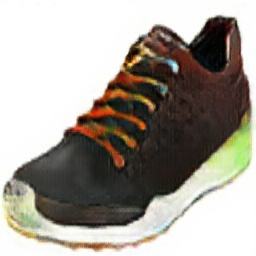}
\includegraphics[width=0.15\linewidth]{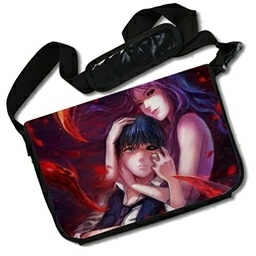}
\includegraphics[width=0.15\linewidth]{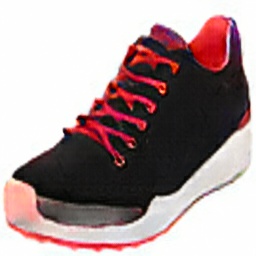}
\\
\includegraphics[width=0.15\linewidth]{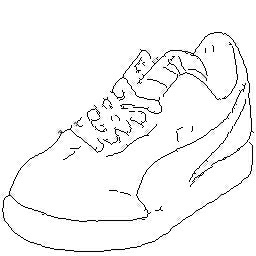}
\includegraphics[width=0.15\linewidth]{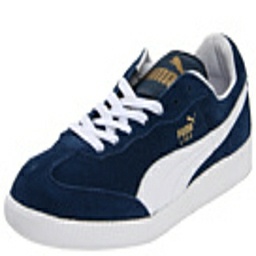}
\includegraphics[width=0.15\linewidth]{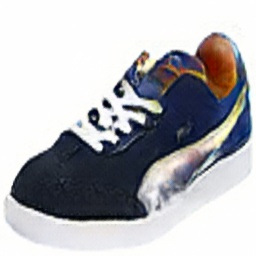}
\includegraphics[width=0.15\linewidth]{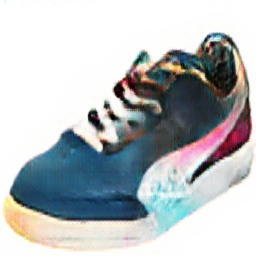}
\includegraphics[width=0.15\linewidth]{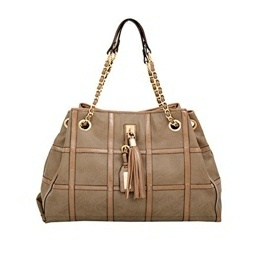}
\includegraphics[width=0.15\linewidth]{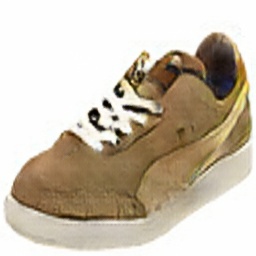}
\\
\includegraphics[width=0.15\linewidth]{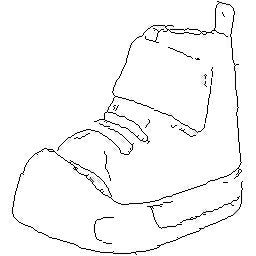}
\includegraphics[width=0.15\linewidth]{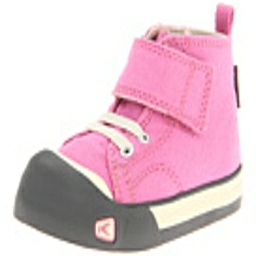}
\includegraphics[width=0.15\linewidth]{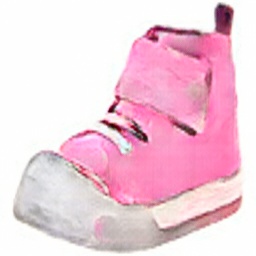}
\includegraphics[width=0.15\linewidth]{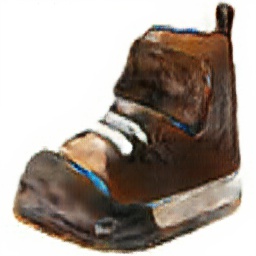}
\includegraphics[width=0.15\linewidth]{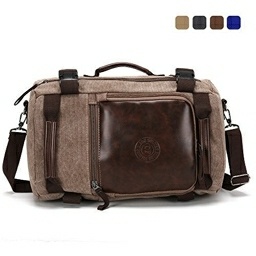}
\includegraphics[width=0.15\linewidth]{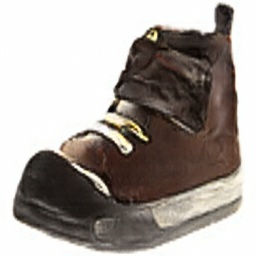}
\\
\includegraphics[width=0.15\linewidth]{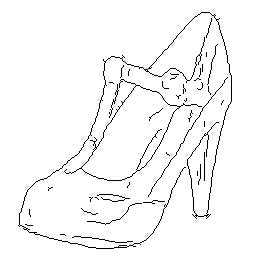}
\includegraphics[width=0.15\linewidth]{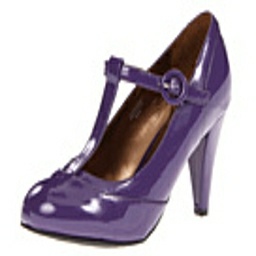}
\includegraphics[width=0.15\linewidth]{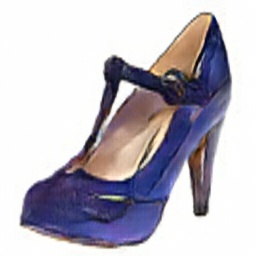}
\includegraphics[width=0.15\linewidth]{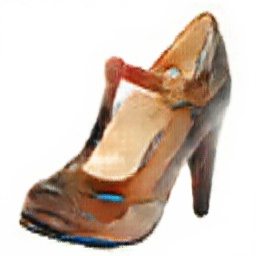}
\includegraphics[width=0.15\linewidth]{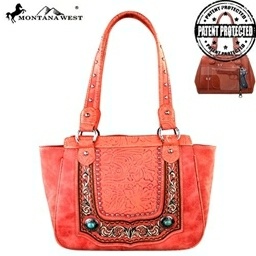}
\includegraphics[width=0.15\linewidth]{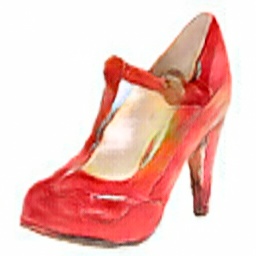}
\\
\includegraphics[width=0.15\linewidth]{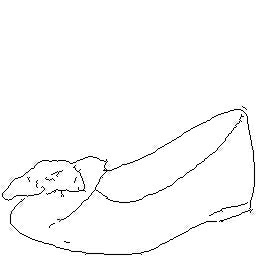}
\includegraphics[width=0.15\linewidth]{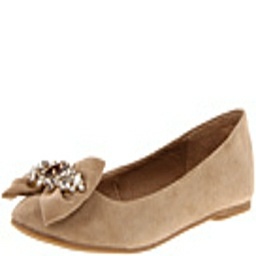}
\includegraphics[width=0.15\linewidth]{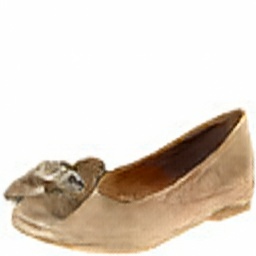}
\includegraphics[width=0.15\linewidth]{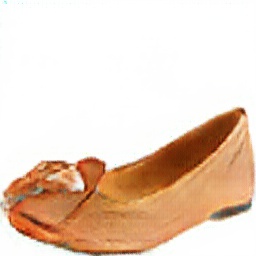}
\includegraphics[width=0.15\linewidth]{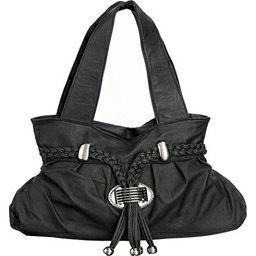}
\includegraphics[width=0.15\linewidth]{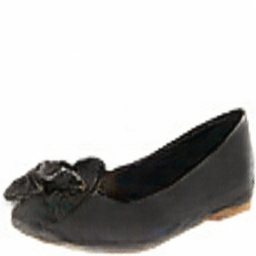}
\\
\includegraphics[width=0.15\linewidth]{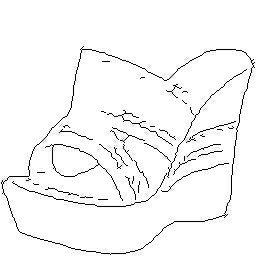}
\includegraphics[width=0.15\linewidth]{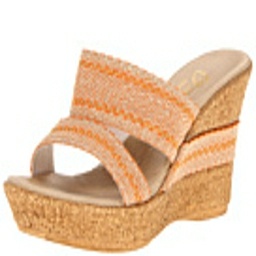}
\includegraphics[width=0.15\linewidth]{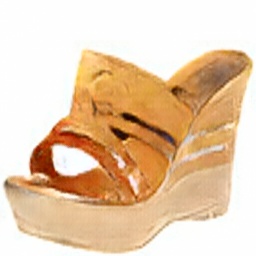}
\includegraphics[width=0.15\linewidth]{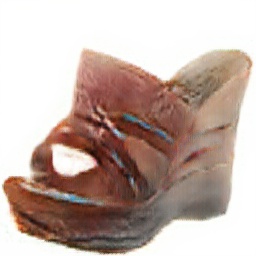}
\includegraphics[width=0.15\linewidth]{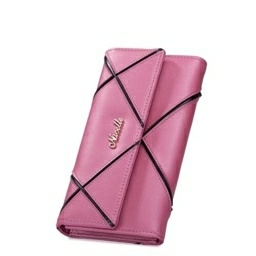}
\includegraphics[width=0.15\linewidth]{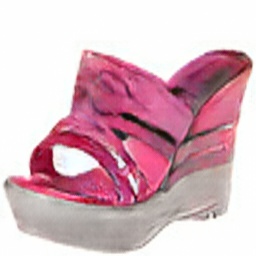}
\\
\includegraphics[width=0.15\linewidth]{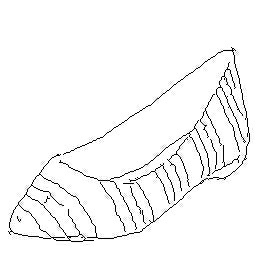}
\includegraphics[width=0.15\linewidth]{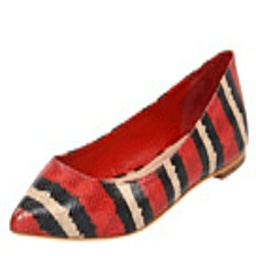}
\includegraphics[width=0.15\linewidth]{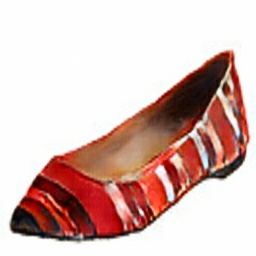}
\includegraphics[width=0.15\linewidth]{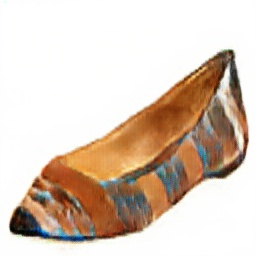}
\includegraphics[width=0.15\linewidth]{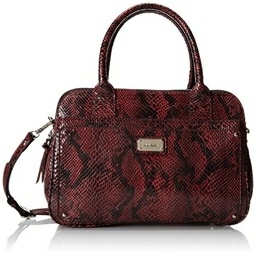}
\includegraphics[width=0.15\linewidth]{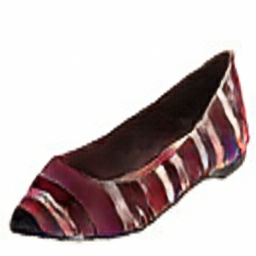}
\end{tabular}
\end{center}
\caption{
Visual results of the edges$\rightarrow$shoe problem.
The output image is generated with the colors of either the source image (col.~3) or a different color reference image (col.~6). For comparison, Pix2Pix~\cite{IsolaZZE16} results are presented (col.~4). 
\label{fig:shoes_table}}
\end{figure}
\begin{figure}
\begin{center}
\begin{tabular}{ccccc}
{\scriptsize Input} & 
{\scriptsize Source} &
{\scriptsize Output (source)} & 
{\scriptsize Reference} & 
{\scriptsize Output (reference)}
\\
\includegraphics[width=0.15\linewidth]{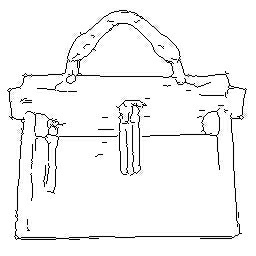}
&\includegraphics[width=0.15\linewidth]{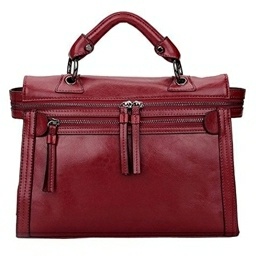}
&\includegraphics[width=0.15\linewidth]{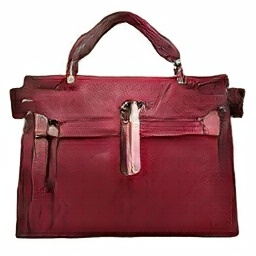}
&\includegraphics[width=0.15\linewidth]{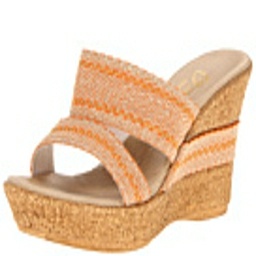}
&\includegraphics[width=0.15\linewidth]{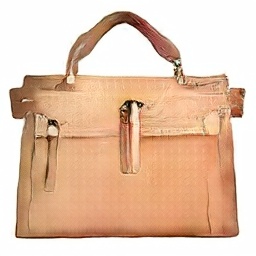}
\\
\includegraphics[width=0.15\linewidth]{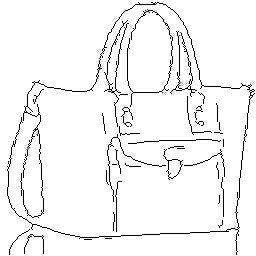}
&\includegraphics[width=0.15\linewidth]{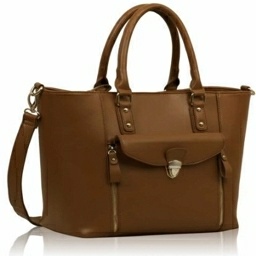}
&\includegraphics[width=0.15\linewidth]{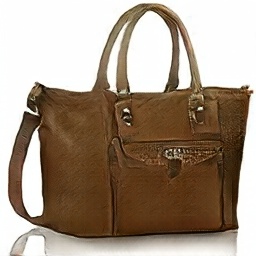}
&\includegraphics[width=0.15\linewidth]{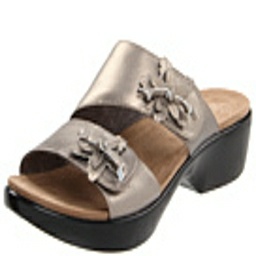}
&\includegraphics[width=0.15\linewidth]{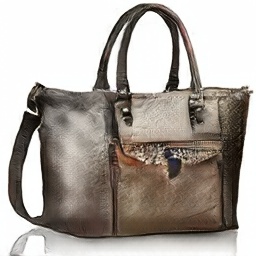}
\\
\includegraphics[width=0.15\linewidth]{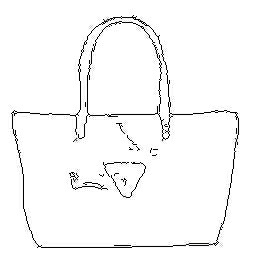}
&\includegraphics[width=0.15\linewidth]{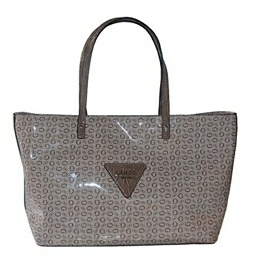}
&\includegraphics[width=0.15\linewidth]{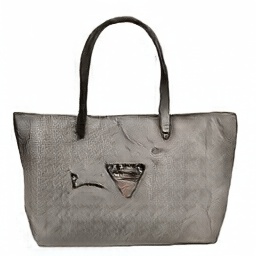}
&\includegraphics[width=0.15\linewidth]{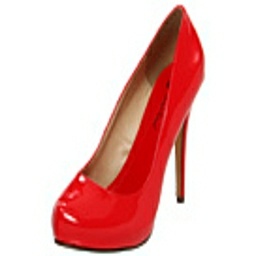}
&\includegraphics[width=0.15\linewidth]{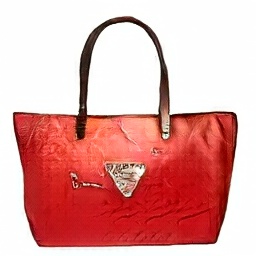}
\\
\includegraphics[width=0.17\linewidth]{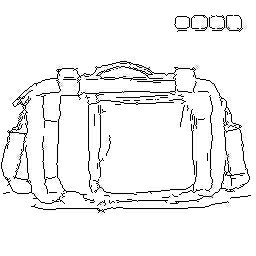}
&\includegraphics[width=0.17\linewidth]{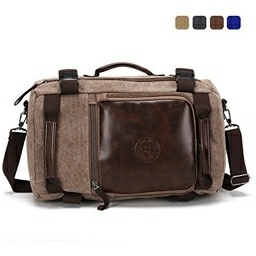}
&\includegraphics[width=0.17\linewidth]{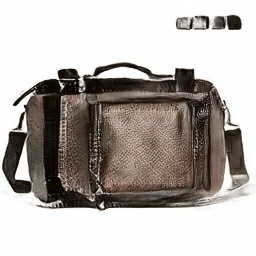}
&\includegraphics[width=0.17\linewidth]{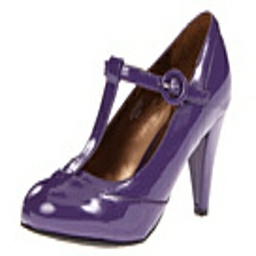}
&\includegraphics[width=0.17\linewidth]{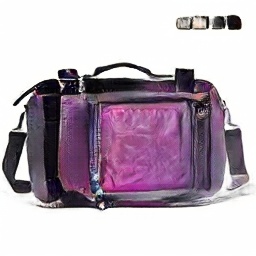}
\\
\includegraphics[width=0.17\linewidth]{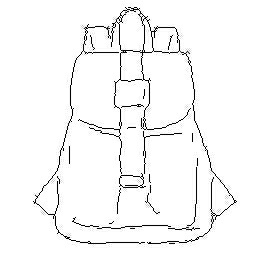}
&\includegraphics[width=0.17\linewidth]{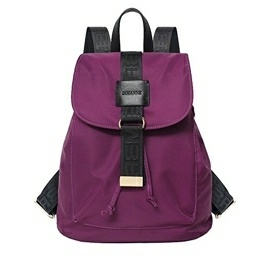}
&\includegraphics[width=0.17\linewidth]{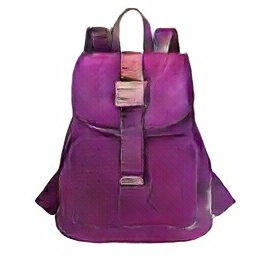}
&\includegraphics[width=0.17\linewidth]{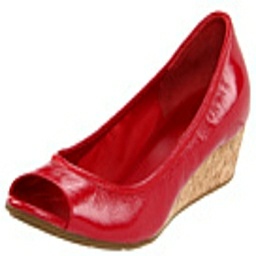}
&\includegraphics[width=0.17\linewidth]{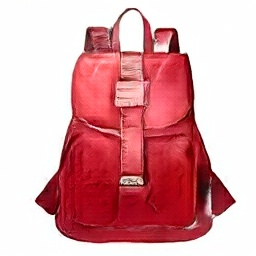}
\\
\includegraphics[width=0.17\linewidth]{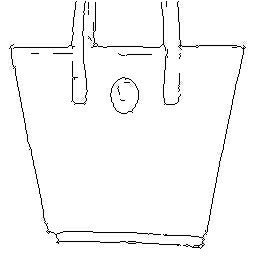}
&\includegraphics[width=0.17\linewidth]{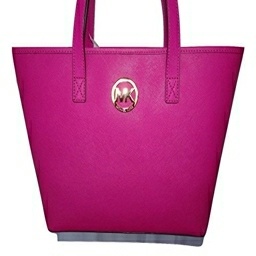}
&\includegraphics[width=0.17\linewidth]{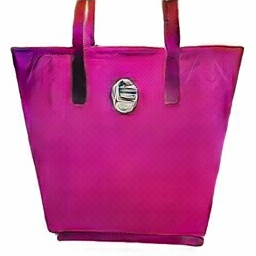}
&\includegraphics[width=0.17\linewidth]{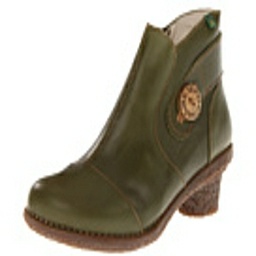}
&\includegraphics[width=0.17\linewidth]{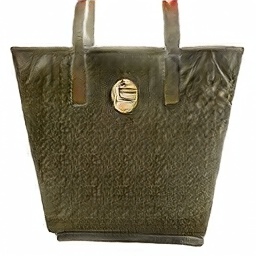}
\\
\includegraphics[width=0.17\linewidth]{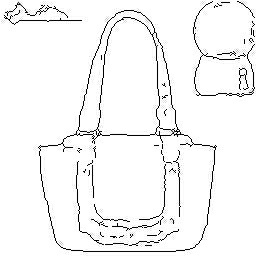}
&\includegraphics[width=0.17\linewidth]{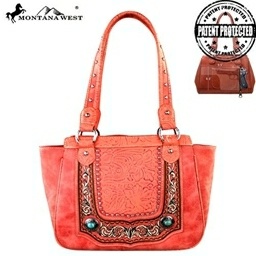}
&\includegraphics[width=0.17\linewidth]{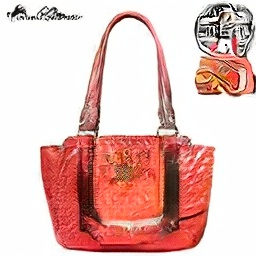}
&\includegraphics[width=0.17\linewidth]{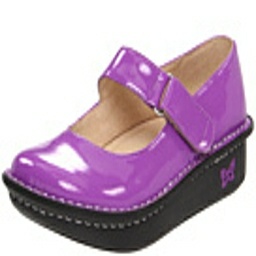}
&\includegraphics[width=0.17\linewidth]{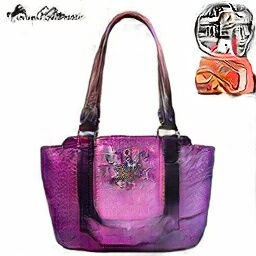}
\\
\includegraphics[width=0.17\linewidth]{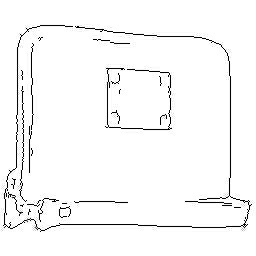}
&\includegraphics[width=0.17\linewidth]{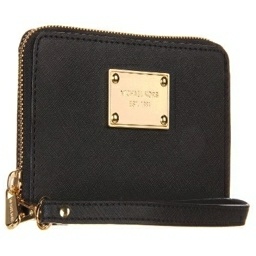}
&\includegraphics[width=0.17\linewidth]{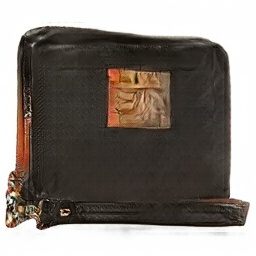}
&\includegraphics[width=0.17\linewidth]{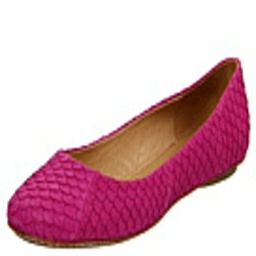}
&\includegraphics[width=0.17\linewidth]{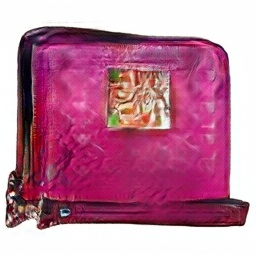}
\end{tabular}
\end{center}
\caption{Visual results of the edges$\rightarrow$bag problem. 
The output image is generated with the colors of either the source image (col.~3) or a different color reference image (col.~5).}
\label{fig:bags}
\end{figure}
\item {\bf Image colorization}
We use the summer/winter Yosemite dataset, prepared by~\cite{zhu2017unpaired} using Flickr API. 
We use train/test splits as in~\cite{zhu2017unpaired}.
During training, the input is a gray-scale image (generated from the source image), randomly selected from the training set of summer and winter images and the output is a colorized image with color distributions of the source (original) image.
During the test phase, we generate synthesized images based on the same gray-scale image yet with different selected color histograms. 
For evaluating our method, we present synthesized images with the color distribution of either the source image or a color reference image.
Results and comparison to CycleGAN are shown in Figure~\ref{fig:colorization}. We note that the results obtained by the CycleGan are much less colourful than the DeepHist results. 

\begin{figure}
\begin{center}
\begin{tabular}{l}
~~~~~~~{\scriptsize Input} 
~~~~~~~~~~~~~{\scriptsize Source 1}
~~~~~~~~~{\scriptsize Output (1)}
~~~~~~~~~{\scriptsize Output (2)}
~~~~{\scriptsize CycleGAN (winter)}
\\
\includegraphics[width=0.19\linewidth]{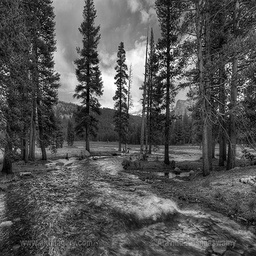}
\includegraphics[width=0.19\linewidth]{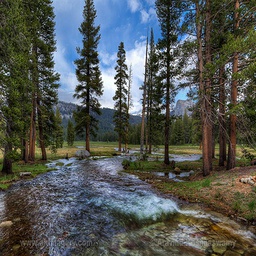}
\includegraphics[width=0.19\linewidth]{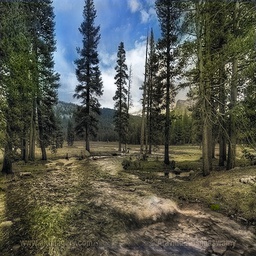}
\includegraphics[width=0.19\linewidth]{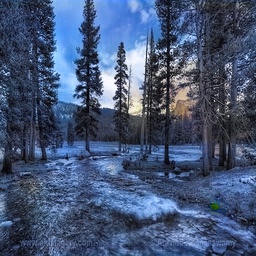}
~\includegraphics[width=0.19\linewidth]{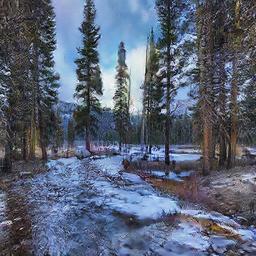}
\\
~~~~~~~{\scriptsize Input} 
~~~~~~~~~~~~~{\scriptsize Source 2}
~~~~~~~~~{\scriptsize Output (2)}
~~~~~~~~~{\scriptsize Output (1)}
~~~~{\scriptsize CycleGAN (summer)}
\\
\includegraphics[width=0.19\linewidth]{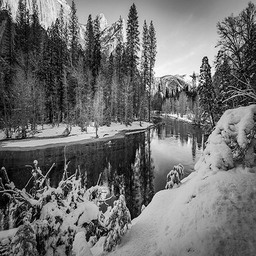}
\includegraphics[width=0.19\linewidth]{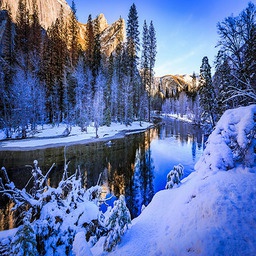}
\includegraphics[width=0.19\linewidth]{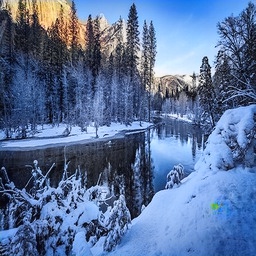}
\includegraphics[width=0.19\linewidth]{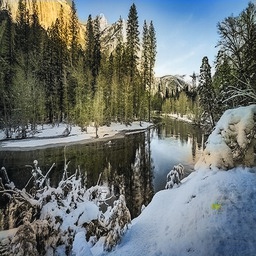}
~\includegraphics[width=0.19\linewidth]{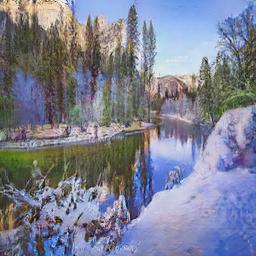}
\\
\midrule
~~~~~~~{\scriptsize Input} 
~~~~~~~~~~~~~{\scriptsize Source 1}
~~~~~~~~~{\scriptsize Output (1)}
~~~~~~~~~{\scriptsize Output (2)}
~~~~{\scriptsize CycleGAN (winter)}
\\
\includegraphics[width=0.19\linewidth]{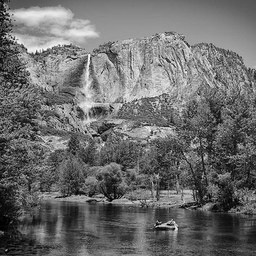}
\includegraphics[width=0.19\linewidth]{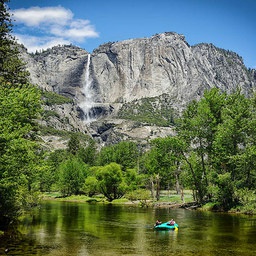}
\includegraphics[width=0.19\linewidth]{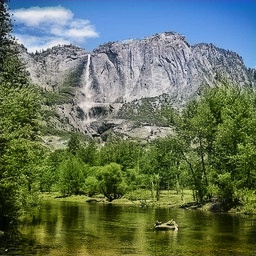}
\includegraphics[width=0.19\linewidth]{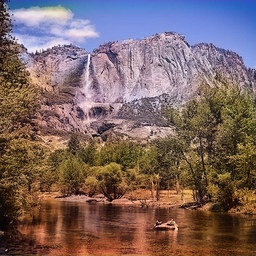}
~\includegraphics[width=0.19\linewidth]{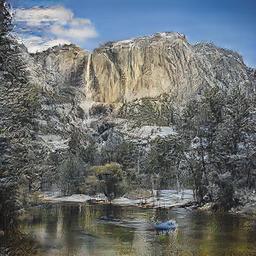}
\\
~~~~~~~{\scriptsize Input} 
~~~~~~~~~~~~~{\scriptsize Source 2}
~~~~~~~~~{\scriptsize Output (2)}
~~~~~~~~~{\scriptsize Output (1)}
~~~~{\scriptsize CycleGAN (summer)}
\\
\includegraphics[width=0.19\linewidth]{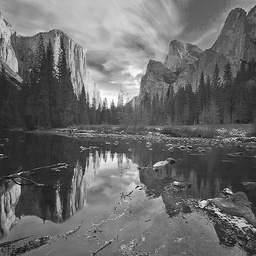}
\includegraphics[width=0.19\linewidth]{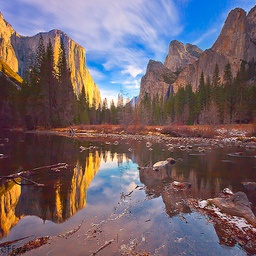}
\includegraphics[width=0.19\linewidth]{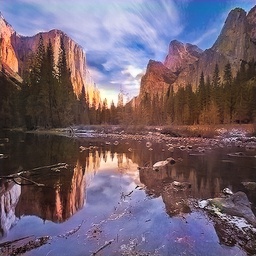}
\includegraphics[width=0.19\linewidth]{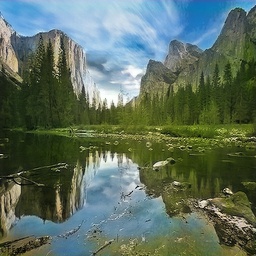}
~\includegraphics[width=0.19\linewidth]{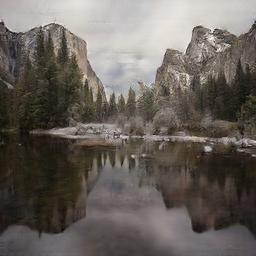}
\\
\midrule
~~~~~~~{\scriptsize Input} 
~~~~~~~~~~~~~{\scriptsize Source 1}
~~~~~~~~~{\scriptsize Output (1)}
~~~~~~~~~{\scriptsize Output (2)}
~~~~{\scriptsize CycleGAN (winter)}
\\
\includegraphics[width=0.19\linewidth]{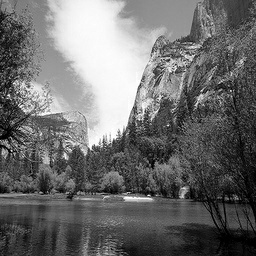}
\includegraphics[width=0.19\linewidth]{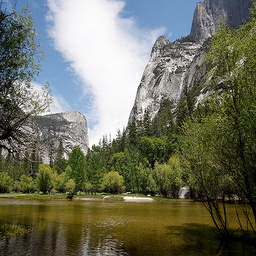}
\includegraphics[width=0.19\linewidth]{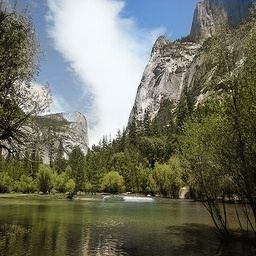}
\includegraphics[width=0.19\linewidth]{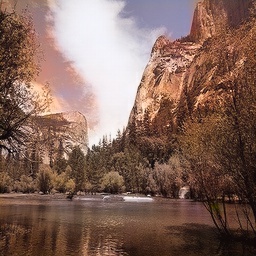}
~\includegraphics[width=0.19\linewidth]{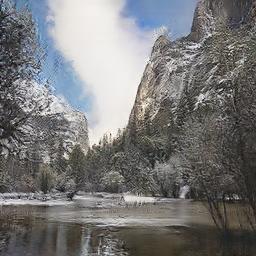}
\\
~~~~~~~{\scriptsize Input} 
~~~~~~~~~~~~~{\scriptsize Source 2}
~~~~~~~~~{\scriptsize Output (2)}
~~~~~~~~~{\scriptsize Output (1)}
~~~~{\scriptsize CycleGAN (summer)}
\\
\includegraphics[width=0.19\linewidth]{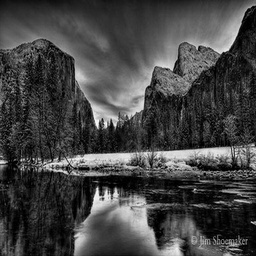}
\includegraphics[width=0.19\linewidth]{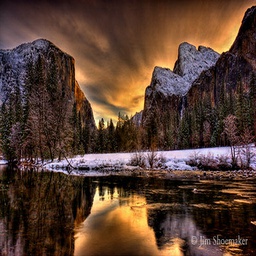}
\includegraphics[width=0.19\linewidth]{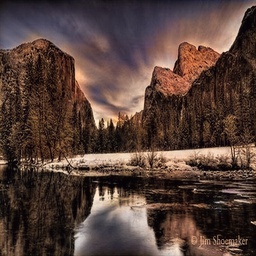}
\includegraphics[width=0.19\linewidth]{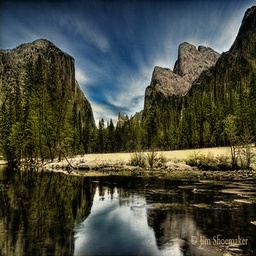}
~\includegraphics[width=0.19\linewidth]{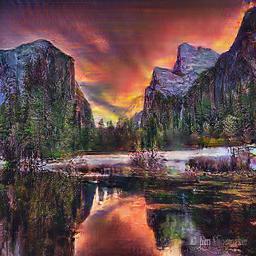}
\end{tabular}
\end{center}
\caption{Visual results of the image colorization problem. 
An input gray-scale image is painted with the colors of either of two source color distributions.
Specifically, output (1) or output (2) refers to the colors of source image 1 or 2, respectively.
Comparison to CycleGAN~\cite{zhu2017unpaired} is presented in column~5.}
\label{fig:colorization}
\end{figure}
\item {\bf Color transfer}
We used the Oxford 102 Category Flower Dataset~\cite{Nilsback08}, which consists of $8189$ images. The dataset was randomly divided  into $7370$ and $819$ images for training and test, respectively.
During training, the input consists of an input and a color reference images that were randomly selected. The aim is to paint the output image in the colors of the reference.
Figure~\ref{fig:flowers} presents color transferred images obtained with and without the MI loss, demonstrating the contribution of the MI loss.
To further justify the use of MI loss we calculated the MI between the input and the output images for all three color channels.
As expected (and desired), the MI between the output and the source is higher using all three DeepHist loss functions rather than without the MI loss. Results are shown in Table~\ref{tbl:MI}. The implication is that the content of the input is better preserved when using the MI loss. This can be also visually observed in Figure~\ref{fig:flowers} when comparing the third and the fourth columns. 

\begin{figure}
\begin{center}
\begin{tabular}{cccc}
{\scriptsize Source}
&{\scriptsize Target}
&{\scriptsize DeepHist}
&{\scriptsize w/o $\mathcal{L}_{\scriptsize\mbox{MI}}$}
\\
\includegraphics[width=0.2\linewidth]{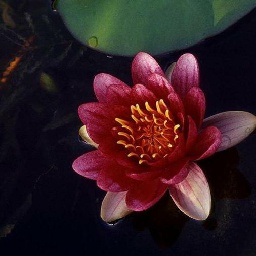}
&\includegraphics[width=0.2\linewidth]{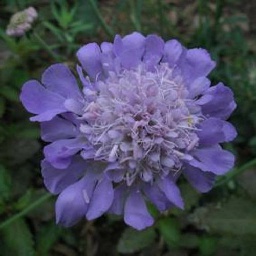}
&\includegraphics[width=0.2\linewidth]{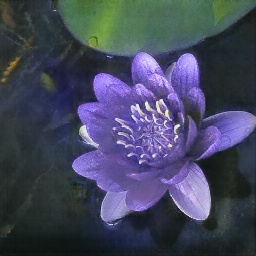}
&\includegraphics[width=0.2\linewidth]{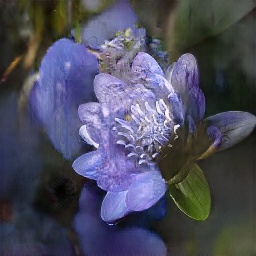}
\\
\includegraphics[width=0.2\linewidth]{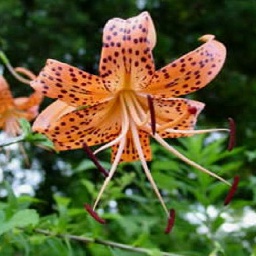}
&\includegraphics[width=0.2\linewidth]{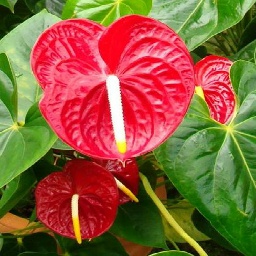}
&\includegraphics[width=0.2\linewidth]{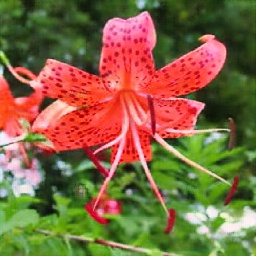}
&\includegraphics[width=0.2\linewidth]{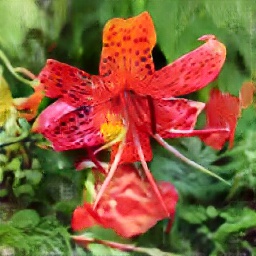}
\\
\includegraphics[width=0.2\linewidth]{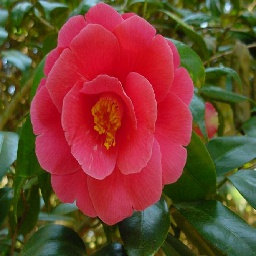}
&\includegraphics[width=0.2\linewidth]{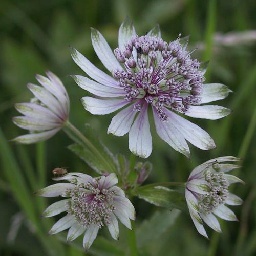}
&\includegraphics[width=0.2\linewidth]{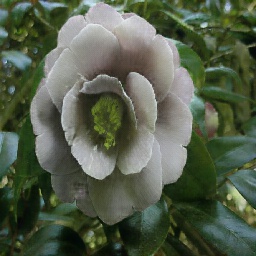}
&\includegraphics[width=0.2\linewidth]{figures/flowers/11_21_42/0_out.jpg}
\\
\includegraphics[width=0.2\linewidth]{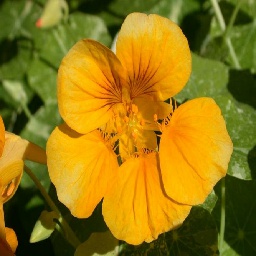}
&\includegraphics[width=0.2\linewidth]{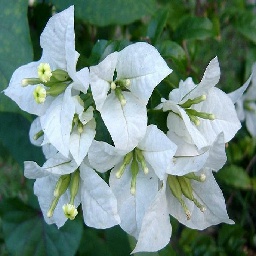}
&\includegraphics[width=0.2\linewidth]{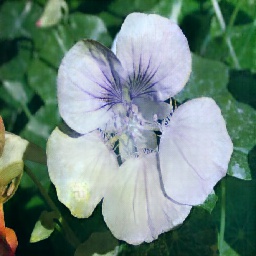}
&\includegraphics[width=0.2\linewidth]{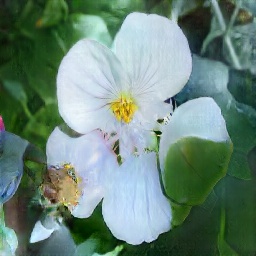}
\\
\includegraphics[width=0.2\linewidth]{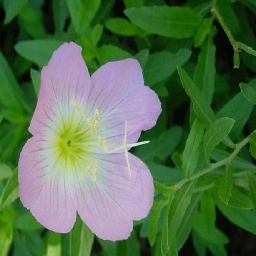}
&\includegraphics[width=0.2\linewidth]{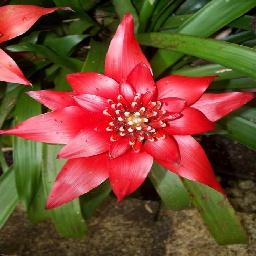}
&\includegraphics[width=0.2\linewidth]{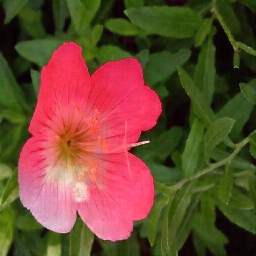}
&\includegraphics[width=0.2\linewidth]{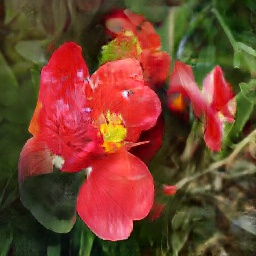}
\end{tabular}
\end{center}
\caption{
Visual color transfer results of the proposed framework compared to the framework trained without the MI loss $\mathcal{L}_{\scriptsize\mbox{MI}}.$\label{fig:flowers}}
\end{figure}

\begin{table}
\begin{center}
\caption{{\bf MI results for the color transfer problem.} Average MI results for each of the three color channels between the generated color transferred images and the respective input images. The comparison is made for the DeepHist framework, using and not using the MI loss.\label{tbl:MI}}  
\begin{tabular}{c|c|c|c}
Loss
 & Y
 & U
 & V
\\
\hline
 DeepHist
 & \textbf{0.178}
 & \textbf{0.141}
 & \textbf{0.167}
\\
\hline
 w/o $\mathcal{L}_{\scriptsize\mbox{MI}}$
 & 0.066
 & 0.035
 & 0.044
\end{tabular}
\end{center}
\end{table} 
\end{enumerate}

\subsection{Perceptual Realism}
\begin{figure}\centering
\includegraphics[width=0.5\linewidth]{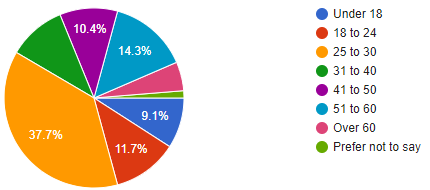}
\caption{Age distribution of the participants in our DeepHist Questionnaire}
\label{fig:age_statistics}
\end{figure}

Addressing the problem of color transfer, the aim is to paint an input image with the colors of a different target image. Note that the desired output image does not exist and therefore we cannot measure the results by quantitative comparison (pixel-to-pixel) of the output image to a ground truth image.
For evaluating the `realism' of our color transfer results we set up a questionnaire for a human observer, in which we presented real (reference) or color transferred (output) images in a random order.
The questionnaire based on our generated images and the true ones can be accessed via~\url{https://forms.gle/NN6HB4Sbr5fDPYo1A}.
Overall, we used $24$ images, of which $12$ were real and $12$ were painted. 
Participants were asked to mark `real' or `fake'. 
Specifically, the following instructions are presented:\\
{\it The following questionnaire shows real pictures of flowers and pictures of flowers that were obtained by painting (changing the colors) of real flower images using a deep learning approach.
Can you tell whether these images are Real or Fake?}\\
We distributed the questionnaire anonymously with the social net (via WhatsApp).
The statistics presented here are based on nearly 100 questionnaire participants, of age groups as shown in Figure~\ref{fig:age_statistics}.
\begin{figure}\centering
\includegraphics[width=0.7\linewidth]{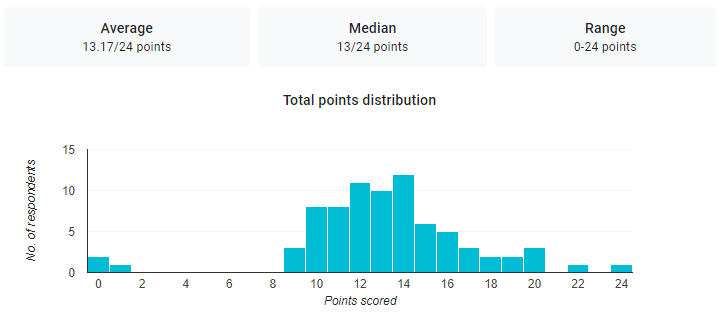}
\caption{The total distribution of the correct results (points) out of 24 questions.}
\label{fig:total_points_distribution}
\end{figure}
\begin{table}
 \caption{Average percentage of participants who marked the target (first row) or the output (second row) as "Real" (first column) or "Fake" (second column). True-Positive (Target, Real), True-Negative (Target, Fake), False-Positive (Output, Real) and False-Negative (Output, Fake) statistics are shown in the table.}
\centering
 \begin{tabular}{c| c c} 
 &Real & Fake \\
 \hline
 Target (real image) & 65.8 & 34.2 \\
 Output (painted image)& 51.9 & 48.1
 \end{tabular}
  \label{tbl:RealFake}
\end{table}\\
The distribution of the number of correct answers is shown in Figure~\ref{fig:total_points_distribution}.
Confusion matrix of the average percentage of participants who marked the target or the output images by either "Real" or "Fake"  is shown in Table~\ref{tbl:RealFake}. 
As shown in the Table, the DeepHist color transfer results misled (on the average) the questionnaire participants on about half of the cases. Moreover, 51.9\% of the synthesized (fake) images were marked as real.

\subsection{Ablation study}
\begin{figure}[h]
\begin{center}
\begin{tabular}{l}
~~~~~{\scriptsize Input}
~~~~~~~~~{\scriptsize Source}
~~~~~~~~{\scriptsize DeepHist}
~~~~~{\scriptsize w/o $\mathcal{L}_{\scriptsize\mbox{EMD}}$}
~~~~~{\scriptsize w/o $\mathcal{L}_{\scriptsize\mbox{MI}}$}
~~~~~{\scriptsize w/o $\mathcal{L}_{\scriptsize\mbox{ADV}}$}
\\
\includegraphics[width=0.15\linewidth]{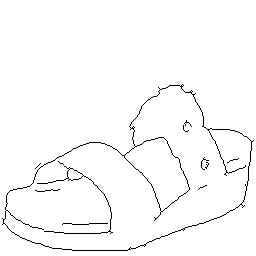}
\includegraphics[width=0.15\linewidth]{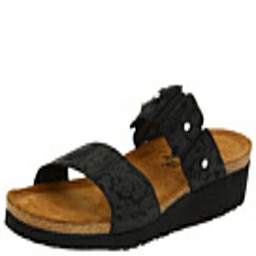}
\includegraphics[width=0.15\linewidth]{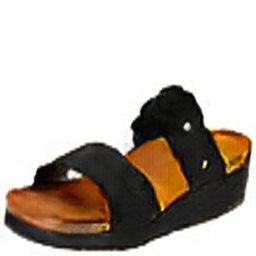}
\includegraphics[width=0.15\linewidth]{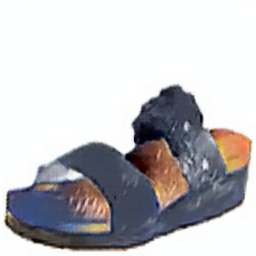}
\includegraphics[width=0.15\linewidth]{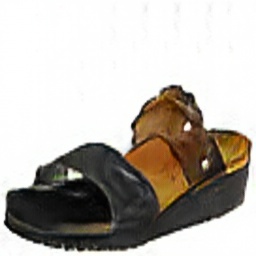}
\includegraphics[width=0.15\linewidth]{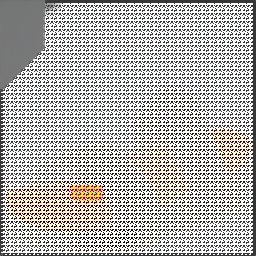}
\\
\includegraphics[width=0.15\linewidth]{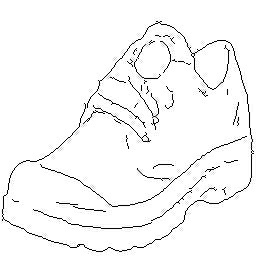}
\includegraphics[width=0.15\linewidth]{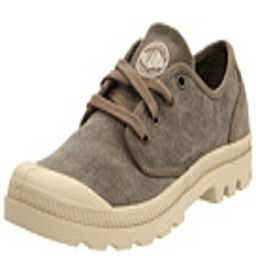}
\includegraphics[width=0.15\linewidth]{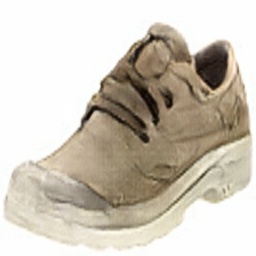}
\includegraphics[width=0.15\linewidth]{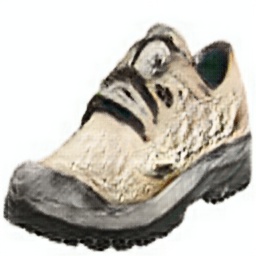}
\includegraphics[width=0.15\linewidth]{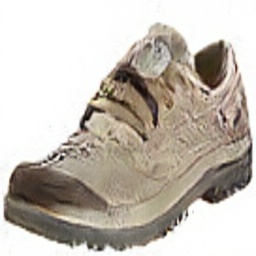}
\includegraphics[width=0.15\linewidth]{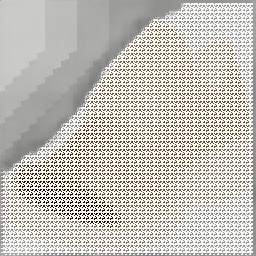}
\\
\includegraphics[width=0.15\linewidth]{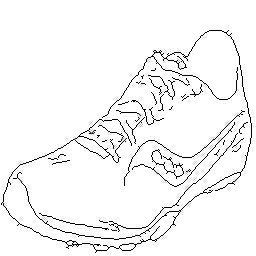}
\includegraphics[width=0.15\linewidth]{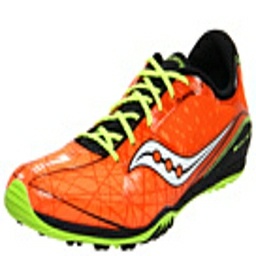}
\includegraphics[width=0.15\linewidth]{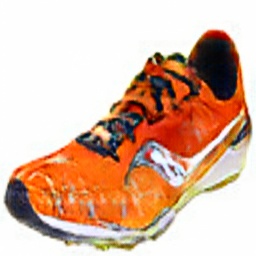}
\includegraphics[width=0.15\linewidth]{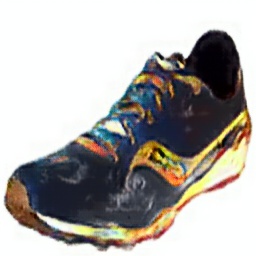}
\includegraphics[width=0.15\linewidth]{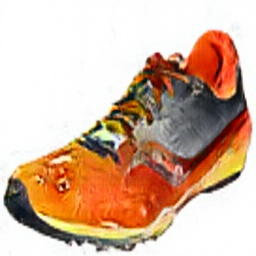}
\includegraphics[width=0.15\linewidth]{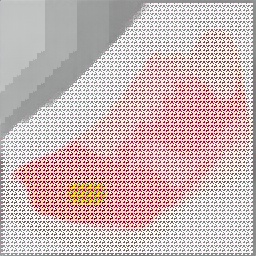}
\end{tabular}
\end{center}
\caption{Ablation study, performance of edges $\rightarrow$ shoes dataset.}
\label{fig:Ablation}
\end{figure}

We run ablation studies to isolate the effect of the EMD
term, the MI term and the GAN term. 
Figure~\ref{fig:Ablation} shows the qualitative effects of these variations on the edges $\rightarrow$ photo problem.
MI and EMD alone (setting $\lambda_{\scriptsize\mbox{ADV}} = 0$ in Eq.~\ref{eq:LossAll}) are not enough to overcome  the "Checkerboard artifact"~\cite{odena2016deconvolution}. 
Using only MI and ADV loss function without the EMD loss (setting $\lambda_{\scriptsize\mbox{EMD}} = 0$ in Eq.~\ref{eq:LossAll}), does not allow the network to adapt the color distribution of the output image to the target color distribution.
Finally, using the ADV and EMD loss functions without the MI loss introduces visual artifacts. The MI loss is important for preserving the content of the  source (regions with the same color). Table~\ref{tbl:edgesMSE} shows that it is also improved the MSE.
We note that in the color-transfer  problem since the discriminator does not have a conditional input, the MI term is essential to preserve the content of the image. Examples are shown in Fig.~\ref{fig:flowers}. 
Table~\ref{tbl:edgesMSE} presents quantitative ablation study results. 
\begin{table}[h]
\begin{center}
\caption{{\bf Ablation study for the edges$\rightarrow$shoe problem}. MSE between the source images and the generated shoe images. The calculation was performed for 200 test images. Specifically we compared the MSE results obtained for the DeepHist methods implemented with all three loss function with respect to the MSE results obtained without either of the loss functions. All image values are in the range $[-1,1]$.\label{tbl:edgesMSE}}
\begin{tabular}{c|c|c|c|c}
& w/o $\mathcal{L}_{\scriptsize\mbox{ADV}}$
& w/o $\mathcal{L}_{\scriptsize\mbox{EMD}}$
& w/o $\mathcal{L}_{\scriptsize\mbox{MI}}$
& DeepHist
\\
\hline
MSE average (std) &.31 (.077)& .053 (.013) & .024(.012)  &0.018(.010)
\end{tabular}
\end{center}
\end{table}

\subsection{Colorfulness}
\begin{figure}[h]
\begin{center}
\begin{tabular}{l}
\hspace{0.17\linewidth}Y 
\hspace{0.3\linewidth}U
\hspace{0.3\linewidth}V
\\
\includegraphics[width=0.32\linewidth]{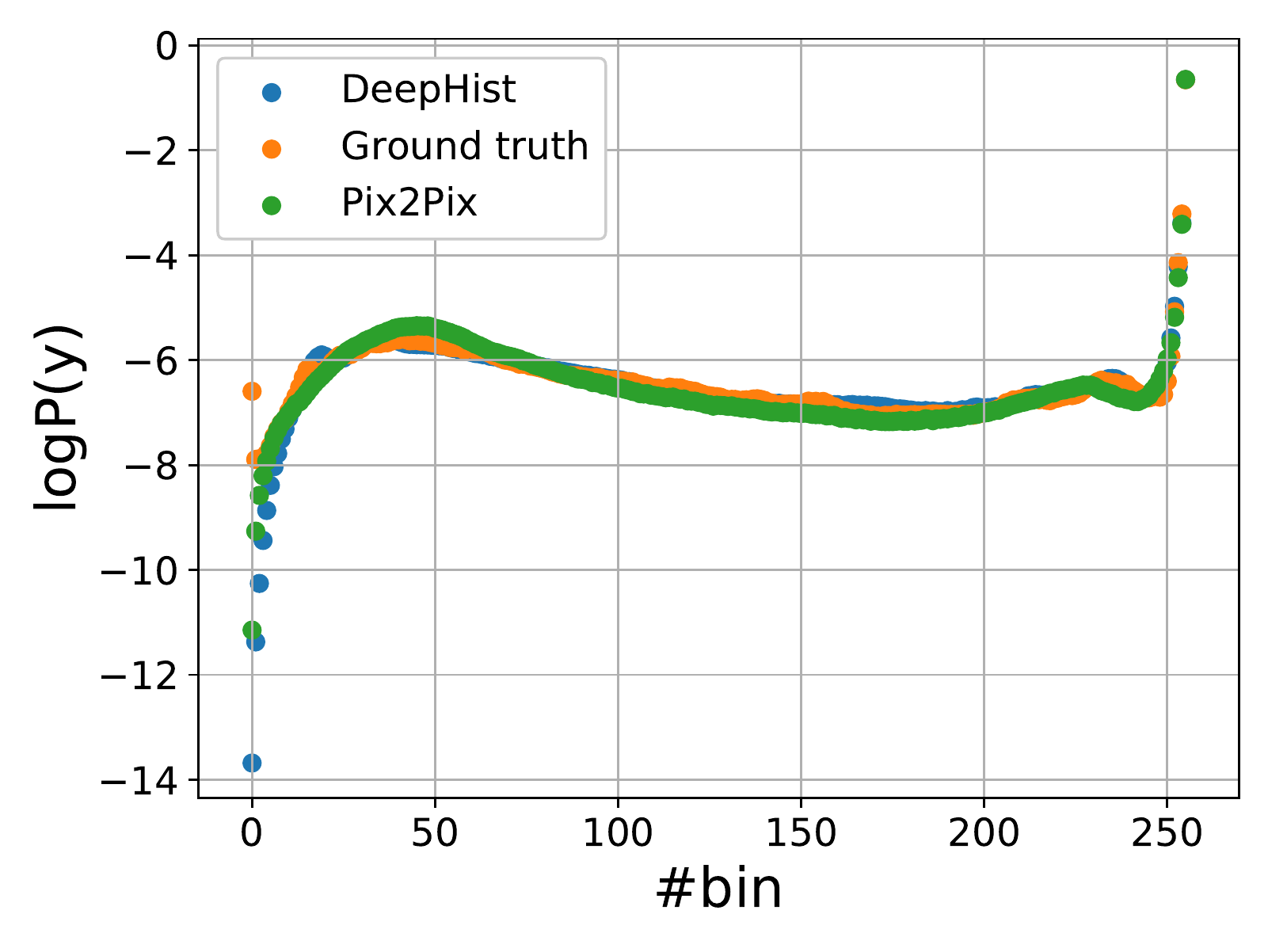}
\includegraphics[width=0.32\linewidth]{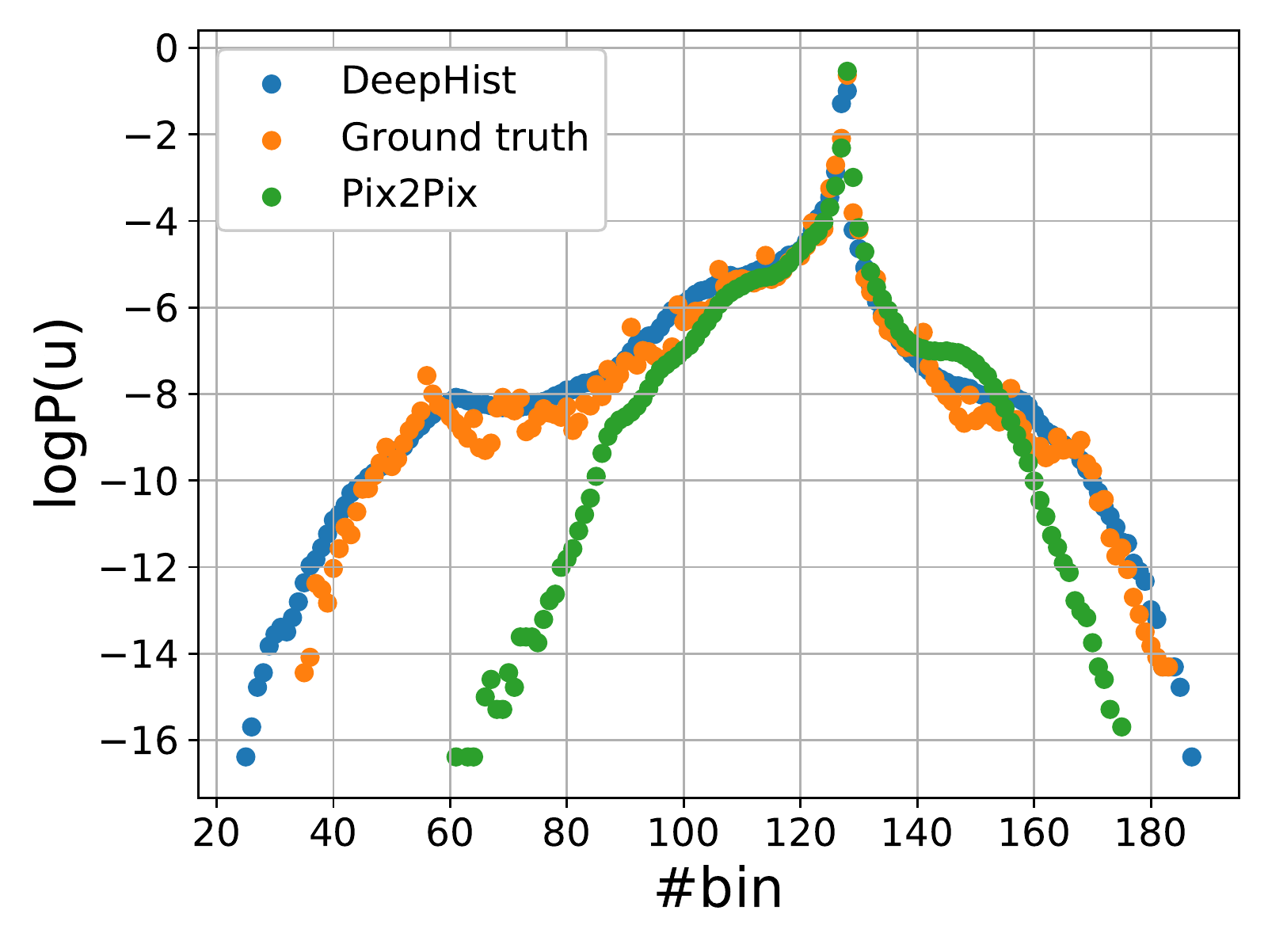}
\includegraphics[width=0.32\linewidth]{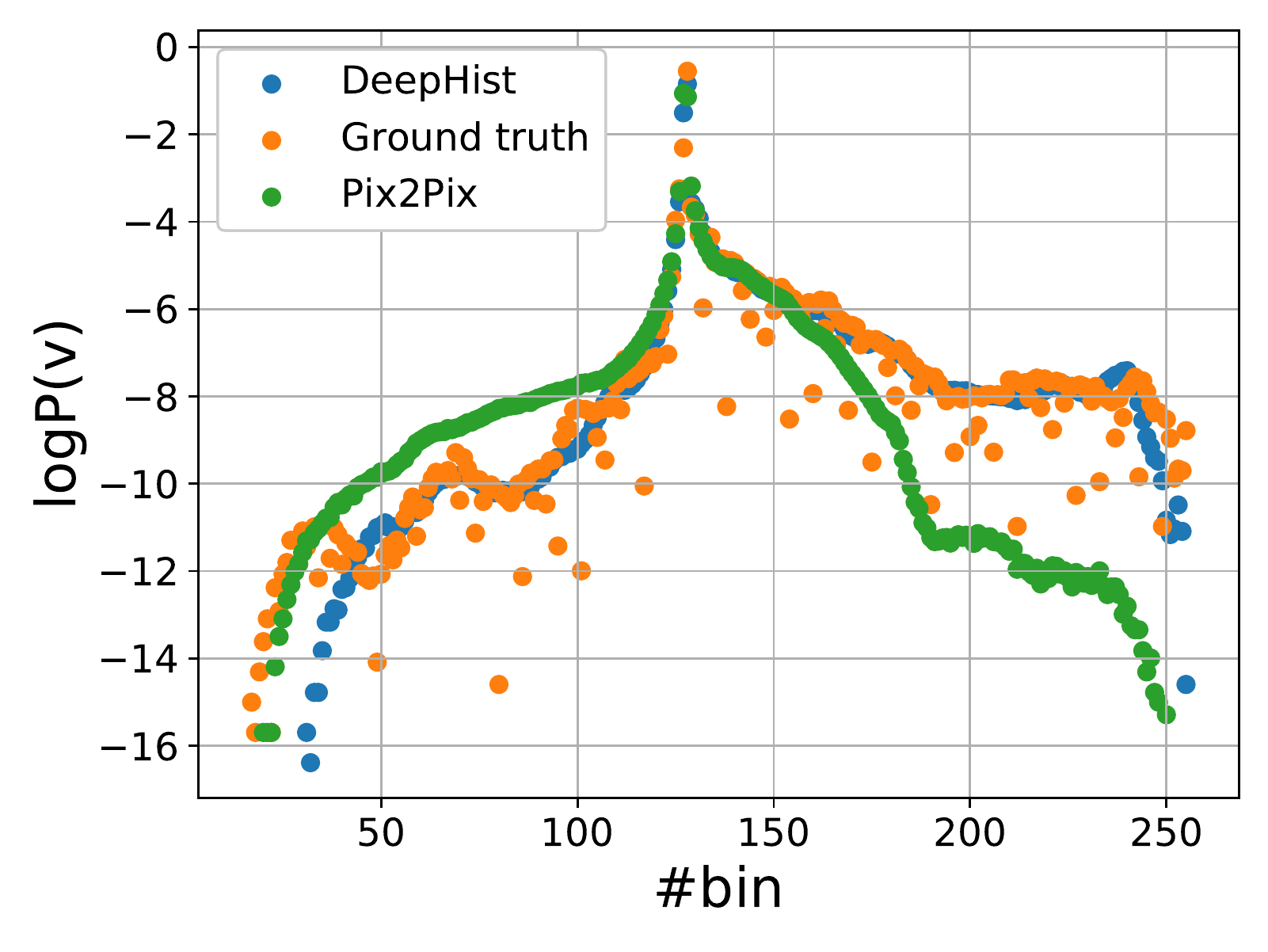}
\end{tabular}
\end{center}
\caption{{\bf Gray level distributions for the output edges$\rightarrow$shoe images.} The plots show the gray-level distributions for the Y, U and V color channels, comparing the proposed DeepHist (blue) with Pix2Pix (green) and the actual color images (orange).\label{fig:color_distribution}}
\end{figure}

As discussed in the pix2pix paper~\cite{IsolaZZE16} the images generated by using L1 loss tend to have grayish or brownish colors when there is an uncertainty regarding to which of several plausible color values a pixel should take.
Specially, L1 will be minimized by choosing the median of the conditional probability density function over possible colors. In~\cite{IsolaZZE16} it was shown that the conditional GAN loss turns the output images more colorful.
In Figure~\ref{fig:color_distribution}, we demonstrate the gray-scale range obtained for each of the YUV color channels in the generated images for the edges$\rightarrow$shoe dataset. 
The plots show the gray-level distributions using the YUV color space for the entire test set, comparing the proposed DeepHist with Pix2Pix and the actual color images. While there are no significant differences for the Y channel, it is apparent that the DeepHist reflects better the gray-level distribution of the actual images for the U and V channels.  

\section{Conclusions}
We presented the DeepHist, a novel deep learning method for image-to-image translation based on the construction of differentiable histograms and histogram-based loss functions. Specifically, intensity-based and MI loss functions are used to encourage intensity similarity to a reference color distribution and structural similarity to the source image.
The adversarial loss is incorporated to constrain the generation of realistic images, making sure, for example, that the leaves and nor the petals will be painted in green.
While the results are promising we believe that the tools we developed can be applicable to other computer vision tasks with slight modifications, e.g., multi-modal image registration or changing illumination.

%
%
\bibliographystyle{splncs04}
\bibliography{egbib}
\end{document}